\documentclass[onecolumn,amsmath,amssymb,12pt,superscriptaddress,nofootinbib,floatfix]{revtex4}
\pdfoutput=1

\usepackage[latin1]{inputenc}
\usepackage[english]{babel}
\usepackage{amssymb}
\usepackage{amsmath}
\usepackage{amsthm}
\usepackage[]{graphicx}
\usepackage[]{subfigure}
\usepackage{tensor}
\usepackage{color}
\usepackage{cancel}
\usepackage{setspace}
\usepackage{fancyhdr}
\usepackage[bookmarks,linktocpage, colorlinks=true, plainpages = false, citecolor = blue,  linkcolor=blue, urlcolor = blue, filecolor = blue]{hyperref}

\def\1p{{(1p)}}

\def\p0{\phi_0}

\def\be{\begin{equation}}
\def\ee{\end{equation}}
\def\beq{\begin{eqnarray}}
\def\eeq{\end{eqnarray}}

\newcommand{\pkt}{\; .}
\newcommand{\kma}{\; ,}

\def\e{{\rm e}}

\begin{document}

\title{Scalar lumps with a horizon}

\author{George Lavrelashvili}
\email[]{george.lavrelashvili@tsu.ge}
\affiliation{Department of Theoretical Physics, A.Razmadze Mathematical Institute \\
	at I.Javakhishvili Tbilisi State University, GE-0177 Tbilisi, Georgia}
\affiliation{Max Planck Institute for Gravitational Physics \\ (Albert Einstein Institute), 14476 Potsdam-Golm, Germany}
\author{Jean-Luc Lehners}
\email[]{jlehners@aei.mpg.de}
\affiliation{Max Planck Institute for Gravitational Physics \\ (Albert Einstein Institute), 14476 Potsdam-Golm, Germany}
\author{Marc Schneider}
\email[]{marc.schneider@aei.mpg.de}
\affiliation{Max Planck Institute for Gravitational Physics \\ (Albert Einstein Institute), 14476 Potsdam-Golm, Germany}

\date{\today}

\begin{abstract}
\vspace{1cm}
\noindent
We study a self-interacting scalar field theory coupled to gravity and are interested in spherically symmetric solutions with a regular origin surrounded by a horizon. For a scalar potential containing a barrier, and using the most general spherically symmetric ansatz, we show that in addition to the known static, oscillating solutions discussed earlier in the literature there exist new classes of solutions which appear in the strong field case. For these solutions the spatial sphere shrinks either beyond the horizon, implying a collapsing universe outside of the cosmological horizon, or it shrinks already inside of the horizon, implying the existence of a black hole surrounding the scalar lump in all directions. Crucial for the existence of all such solutions is the presence of a scalar field potential with a barrier that satisfies the swampland conjectures.
\end{abstract}
\maketitle
%\newpage
%	\tableofcontents
	%\newpage

\section{Introduction}

Scalar fields play an important role in modelling and explaining various phenomena in cosmology: for example, they are essential in many models of
inflation, ekpyrosis, modified gravity or quintessence, for reviews see e.g. \cite{Martin:2013tda,Lehners:2008vx,Clifton:2011jh}. Thus it is of evident interest to study the possible solutions that arise in the presence of scalar fields. For this, the shape of their potential is crucial. As has been extensively discussed in recent years, there may exist significant restrictions (within the framework of the swampland conjectures \cite{Obied:2018sgi,Garg:2018reu,Ooguri:2018wrx}) on scalar potentials stemming from the embedding into quantum gravity. Also, in the presence of horizons no hair theorems have been investigated that put constraints on the existence of nontrivial
scalar field configurations.

In particular, it was shown by Torii et al. \cite{Torii:1999uv} that static spherically symmetric solutions with nontrivial scalar field configurations
in asymptotically de Sitter spacetime cannot exist when the potential is convex. However, for non-convex potentials, nontrivial solutions do exit. The examples given in \cite{Torii:1999uv} were for double well potentials, with the minima located at positive values of the potential. The analysis of \cite{Torii:1999uv} was restricted by the use of Schwarzschild like coordinates, in which the size of the $2$-spheres in the metric increases linearly with the radial coordinate. Meanwhile, experience with Einstein-Yang-Mills solutions with nonzero cosmological constant
shows \cite{Volkov:1996qj} that for some parameter values of the theory Schwarzschild like coordinates cannot be used
and one needs to use a different gauge, in which the size of the $2$-spheres is a (possibly non-monotonic) function of the radial coordinate. We encounter the same situation here, for a model of a self-interacting scalar coupled to gravity. We find that Schwarzschild like coordinates are applicable only in a small range of the parameter space of the theory
and, with a more appropriate gauge, new classes of solutions exist for a vastly increased parameter range. These new solutions have a regular origin, which may be identified as the centre of a scalar lump, and develop a horizon at some radius. The transverse $2$-spheres can either grow monotonically or turn around at some distance from the origin. This turnaround can occur either beyond the horizon or even before the horizon is reached. In all cases, the turnaround is associated with a non-trivial oscillation of the scalar field across a potential barrier, and thus it is appropriate to think of the size of the lump as extending to just beyond the turnaround radius. Beyond the turnaround the $2$-sphere radii keep shrinking until a spacetime singularity is reached. The singularity is always located behind the horizon.

In \cite{Torii:1999uv} an inequality on the parameters of the potential was derived, and it was conjectured that for potentials satisfying the inequality, scalar lump solutions with an asymptotic de Sitter region always exist. As alluded to before, we will demonstrate that such solutions only make up a small part of the parameter space satisfying the inequality just mentioned. The largest part consists of the new solutions in which the transverse space starts shrinking again beyond an extremal radius. At least in a heuristic sense, the new turnaround solutions are therefore generic, provided the potential contains an appropriate barrier. In this context it is fitting to remark that the Higgs potential contains at least one (probably two) barriers, offering the prospect that our solutions may have some cosmological relevance.

The plan of this paper is as follows: we present our model in section \ref{sec:model} and discuss a number of its properties that are of relevance to us in section \ref{sec:AnCon}. The solutions themselves have to be found numerically; we present these solutions in section \ref{sec:results}, and elaborate on their causal structure in section \ref{sec:causal}. Finally, we end with a discussion of the features and possible applications of the solutions in section \ref{sec:discussion}.

%%%%%%%%%%%%%%%%%%%%%%%%%%%%%%%%%%%%%%%%%%%%%%%%%

%%%%%%%%%%%%%%%%%%%%%%%%%%%%%%%%%%%%%%%%%%%%%%%%%

\section{Ansatz and field equations} \label{sec:model}

Let us consider a self-interacting scalar field theory minimally coupled to gravity, with action
\begin{align}
S=\int d^4 x \sqrt{-g} \left[\frac{1}{2\kappa}{\mathcal R}- \frac{1}{2}g^{\mu\nu}\partial_\mu\varphi\partial_\nu\varphi - V(\varphi) \right] \kma
\end{align}
where $\kappa = 8 \pi G$. Later we will specialise to a double well potential, but for now we will leave the potential general. In what follows we will assume spherical symmetry.
A general static, spherically symmetric metric can be written as \cite{Landau}
\begin{align}
ds^2= -f(r) \e^{2\delta(r)} dt^2 + \frac{dr^2}{f(r)} +R^2(r) d\Omega^2_2 \kma \label{metricansatz}
\end{align}
where $f, \delta$ and $R$ are functions of the radial coordinate $r$ and $d\Omega^2_2=d\theta^2+\sin^2(\theta) d\phi^2$
is the metric of the unit $2$-sphere. Across a horizon ($f(r)=0$) the ``radial'' coordinate $r$ becomes time-like and $t$ in turn becomes space-like, as is familiar from black hole solutions.
Within the ansatz \eqref{metricansatz} we are still free to perform coordinate transformation which respect spherical symmetry.
This allows e.g. the common choice $R(r) \equiv r$, which we will refer to as Schwarzschild gauge,
\begin{align}\label{eq:interval S gauge}
ds^2= -f(r) \e^{2\delta(r)} dt^2 + \frac{dr^2}{f(r)} +r^2 d\Omega^2_2 \,.
\end{align}
Note that it is only possible to choose this gauge if the size of the 2-spheres $R(r)$ is a monotonic function of $r$, i.e. $dR/dr\neq 0$.
If however a solution develops a turning point $dR/dr = 0$ and $R$ becomes multi-valued, Schwarzschild gauge is not appropriate anymore
and instead we can use e.g. $\delta \equiv 0$ gauge,
\begin{align} \label{eq:interval general gauge}
ds^2= -f(r) dt^2 + \frac{dr^2}{f(r)} +R^2(r) d\Omega^2_2 \pkt
\end{align}
The field equations in this gauge are given by
\begin{align}
\varphi'' &= -(2\frac{R'}{R}+\frac{f'}{f}) \varphi'+ \frac{1}{f}\frac{\partial V}{\partial \varphi} \kma\\
f'' &= -2 \frac{R'}{R} f' - 2 \kappa V \kma \\
R'' &= - \frac{\kappa}{2} R \varphi'^2 \kma \label{eq:R}
\end{align}
while the constraint equation reads
\begin{align}
f \frac{R'^2}{R^2 } + f' \frac{R'}{R} = \frac{1}{R^2 }+ \kappa (\frac{1}{2}f \varphi'^2 -V) \kma \label{eq:constraint}
\end{align}
with primes denoting derivatives w.r.t. $r$.

In this article we will be interested in solutions with a regular origin at $R=0,$ i.e. we will assume that $f(0)\neq 0$. By re-scaling both $f(r)$ and our coordinates we may choose without loss of generality $f(0)=1.$ Then we can expand the solution close to a putative regular origin, thus obtaining the first few terms of a Taylor series,
\begin{align}
\varphi(r) &= \varphi_0 + \frac{V'(\varphi_0)}{6} r^2 + V'(\varphi_0)(\frac{\kappa V(\varphi_0)}{36}+ \frac{V''(\varphi_0)}{120} ) r^4 +O(r^6) \kma \label{eq:phiTaylorZeroGG}\\
f(r) &= 1- \frac{\kappa V(\varphi_0)}{3} r^2 - \frac{ \kappa V'(\varphi_0)^2}{60} r^4
+ O(r^6) \kma \label{eq:fTaylorZeroGG}\\
R(r) &= r -\frac{ \kappa V'(\varphi_0)^2}{360}  r^5
+ O(r^7) \kma \label{eq:RTaylorZeroGG}
\end{align}
where $\varphi_0$ is a free parameter.

As we will demonstrate in what follows, when the potential is positive solutions develop a horizon at some $r=r_h$, where $f(r_h)=0.$  Close to the  horizon one finds a 3-parameter family of regular solutions, with expansion
\begin{align}
\varphi (\rho) &= \varphi_h + \frac{V'(\varphi_h)}{f'_{h}} \rho
+ \frac{V'(\varphi_h) [2 \kappa V(\varphi_h) +V''(\varphi_h)]}{4 {f'_{h}}^2}\rho^2 + O(\rho^3) \kma \label{eq:phiTaylorH} \\
f(\rho) &= f'_{h} \rho - \frac{1}{R_h^2}\rho^2 + O(\rho^3) \kma \label{eq:fTaylorH}\\
R(\rho) &= R_{h} + \frac{1-\kappa R_{h}^2 V(\varphi_h) }{f'_{h} R_{h}} \rho
- \frac{\kappa R_{h} V'(\varphi_h)^2}{4 {f'_{h}}^2}   \rho^2 + O(\rho^3) \kma \label{eq:RTaylorH}
\end{align}
where $\rho=r-r_{h}$ and $f'_{h},\varphi_{h}$ and $R_{h}$ are the three free parameters.

%%%%%%%%%%%%%%%%%%%%%%%%%%%%%%%%%%%%%%%%%%%%%%%%%

%%%%%%%%%%%%%%%%%%%%%%%%%%%%%%%%%%%%%%%%%%%%%%%%%

\section{Analytic considerations}\label{sec:AnCon}

\begin{figure}[h]
	\centering
	\includegraphics[width=0.4\textwidth]{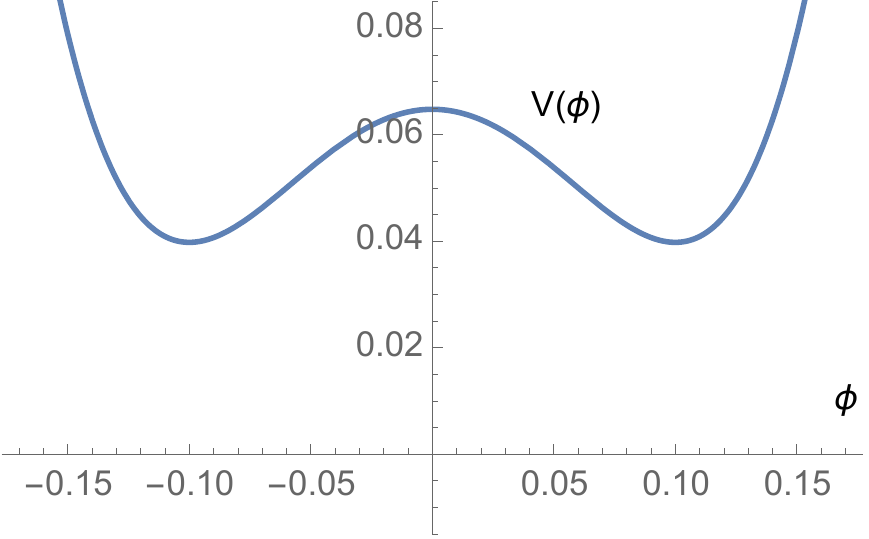}
	\caption{The potential used for the numerical solutions. For simplicity we have chosen the potential to be a symmetric double well, of the form $V(\varphi) = \frac{g}{4} (\varphi^2- v^2)^2 +  \frac{\Lambda}{ \kappa}.$ In the figure the parameters are $\Lambda=1, \kappa = 8\pi, g = 1000, v=1/10.$}
	\label{fig:pot}
\end{figure}

In order to be able to compare our results with \cite{Torii:1999uv}
we will explore a quartic scalar field potential
\be
V(\varphi) = \frac{g}{4} (\varphi^2- v^2)^2 +  \frac{\Lambda}{ \kappa} \pkt
\ee
A simple rescaling of the coordinates $t\to \sqrt{\Lambda}t, r\to \sqrt{\Lambda}r, R\to \sqrt{\Lambda}R$  and
of the scalar field potential $V \to V/\Lambda$ shows that solutions depend only on the ratio
$ \lambda = \frac{g}{\Lambda},$ plus on the value of $v$ (cf. \cite{Torii:1999uv}).
A graph of this potential is provided in Fig.~\ref{fig:pot}.

When the scalar field is non-dynamical one finds the usual de Sitter solution, assuming $\Lambda>0$. Namely when $\varphi = \pm v$, one finds
\be
R(r) = r \kma~~~~ f(r) = 1 - \frac{\Lambda}{3} r^2 \pkt \label{dS}
\ee
and when the scalar field sits on top of the potential, $\varphi \equiv 0$,
the effective cosmological constant is $\Lambda$ plus the height of the potential barrier,
\be
\Lambda_{eff} = \Lambda + \frac{\kappa g v^4}{4} \pkt
\ee

In \cite{Torii:1999uv} numerical solutions were presented in which the scalar field interpolates across the potential barrier. In particular it was shown that the re-scaled coupling constant $\lambda$ needs to be large enough (together with a condition on the allowed range of $v$) in order for such oscillating solutions to exist. More precisely, it was shown that the existence of solutions depends on the desired number $n$ of scalar field interpolations. For $n$ interpolations, an analytic argument was presented showing that $\lambda$ needs to be larger than the critical value
\begin{align}
\lambda_{cr}=\frac{2n(2n+3)}{3v_{cr}^2-4\pi n(2n+3)v_{cr}^4}\,. \label{lambdacritical}
\end{align}
Our numerical investigations have confirmed the existence of these solutions, but also show (as will be demonstrated explicitly below) that for a larger critical value $\lambda^*(v)$ the function
$R(r)$ becomes flat and solutions approach the well known Nariai \cite{Nariai1951} solution at large $r,$
\be
\varphi=\pm v \kma R(r)= \frac{1}{\sqrt{\Lambda}} \kma f(r)= 1 -\Lambda r^2 \pkt
\ee
In this limit the spacetime has become a direct product of $dS^2$ and $S^2,$  and the metric can also be written as
\begin{align}
ds^2 = \frac{1}{\Lambda} \left(-d\tilde{t}^2 + \cosh^2(\tilde{t})dx^2 + d\Omega^2_2 \right) \pkt \label{Nariai}
\end{align}
There are some curious features to this limit, in particular the fact that the spheres are of constant radius everywhere, which appears as a discontinuous feature of the limit compared to the standard de Sitter metric \eqref{dS}. This characteristic suggests that it may not always be appropriate to choose the Schwarzschild gauge where $R(r)=r.$ In the presence of a scalar field this function may turn around and cease to be monotonic. This is implied very clearly by the equation of motion for $R,$ which we repeat here for convenience
\begin{align}
R''=-\frac{\kappa}{2}R\phi^{\prime 2}\,.
\end{align}
A non-trivial scalar field may thus induce a turn in $R,$ and we can see that a non-trivial scalar field profile may achieve a turn not simply in one go, but from successive contributions from regions where the scalar gradient is substantial. This reasoning suggests that new classes of solutions ought to exist at sufficiently large $\lambda,$ in which $R$ turns around. The above equation also implies that although $R$ may take a maximum value, a minimum (and thus also the presence of more than one maximum) is excluded. We are now ready to present our numerical results which confirm these expectations.

%%%%%%%%%%%%%%%%%%%%%%%%%%%%%%%%%%%%%%%%%%%%%%%%%

%%%%%%%%%%%%%%%%%%%%%%%%%%%%%%%%%%%%%%%%%%%%%%%%%

\section{Numerical results} \label{sec:results}

We will be working in $G=1$ units. Our strategy is as follows.
As we already mentioned, solutions with a regular origin behave as Eqs.~(\ref{eq:phiTaylorZeroGG}, \ref{eq:fTaylorZeroGG}, \ref{eq:RTaylorZeroGG}) and regular solutions close to a horizon behave as Eqs.~(\ref{eq:phiTaylorH}, \ref{eq:fTaylorH}, \ref{eq:RTaylorH}).
Having this in mind we numerically integrate the equations of motion from the origin, $r=\epsilon$, and independently from the horizon, $r=r_h-\epsilon$,
with a small $\epsilon$ (typically of order of $10^{-5}$). We then match the solutions at some convenient intermediate point $r=r_*,$ chosen for example such that from both sides the metric function $f$ has reached a specified reference value $f(r_*)=\bar{f}.$ Since the equations of motion are three second order differential equations, we have six functions to start with.
Taking into account the constraint Eq.~(\ref{eq:constraint}) the number of functions to be matched is reduced to five. Moreover, our matching is done such that $f(r)$ is automatically continuous at the matching point. Thus there remain four matching conditions, and accordingly we can adjust the four independent parameters  $\varphi_0, \varphi_h, f'_h, R_h$ using a Newtonian algorithm in order to obtain smooth solutions. In our numerics, we have optimised the solutions such that the initial conditions are accurate to eight significant digits.

\begin{figure}[h]
	\centering
	\includegraphics[width=0.3\textwidth]{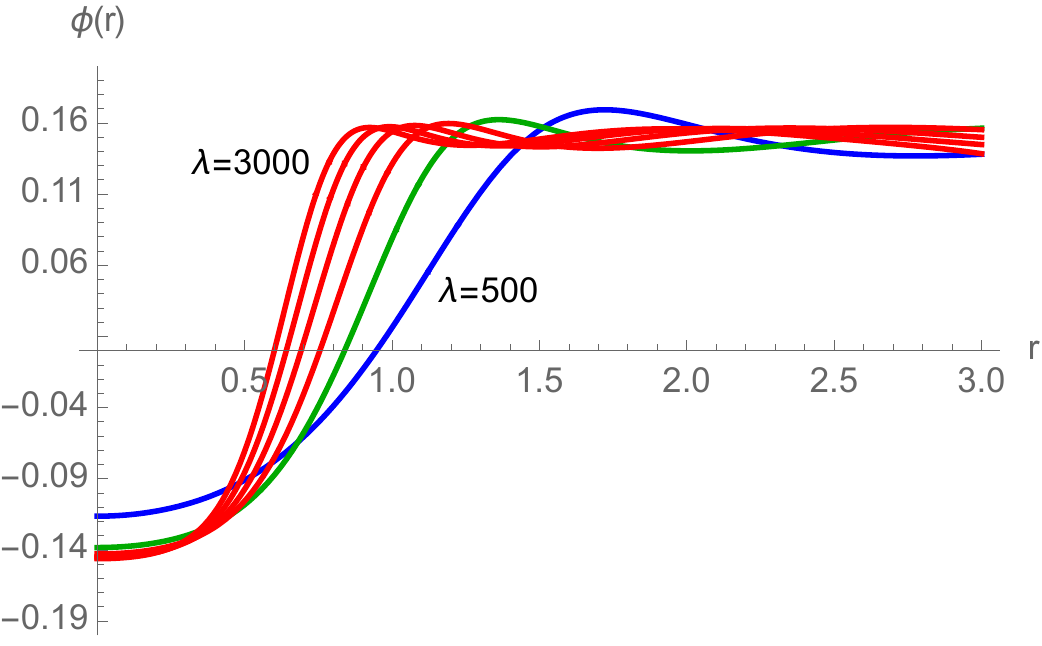}
	\includegraphics[width=0.3\textwidth]{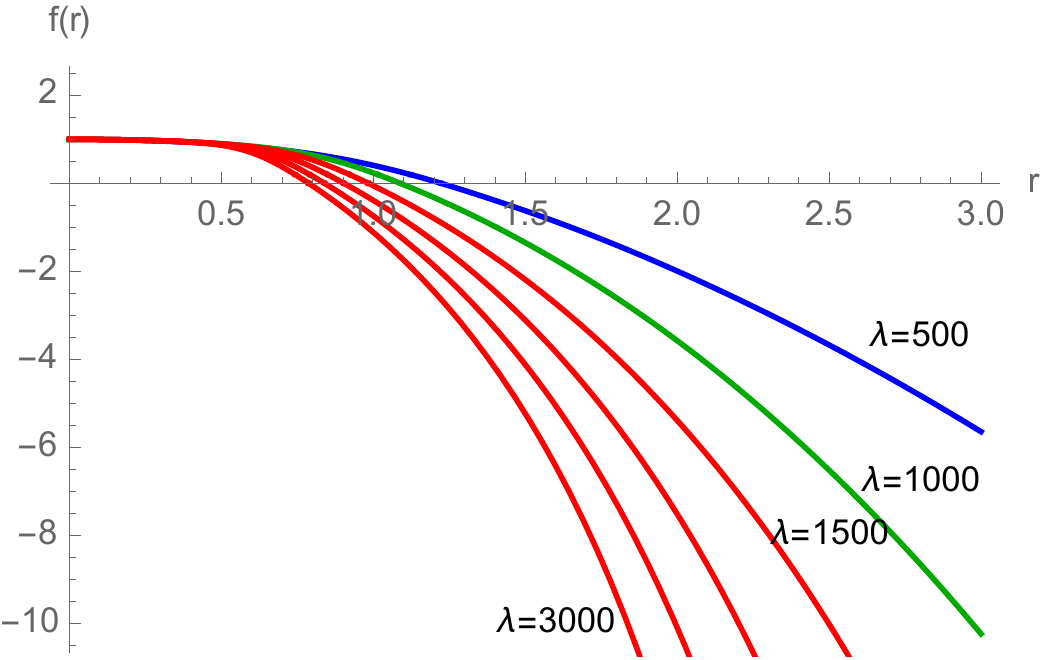}
	\includegraphics[width=0.3\textwidth]{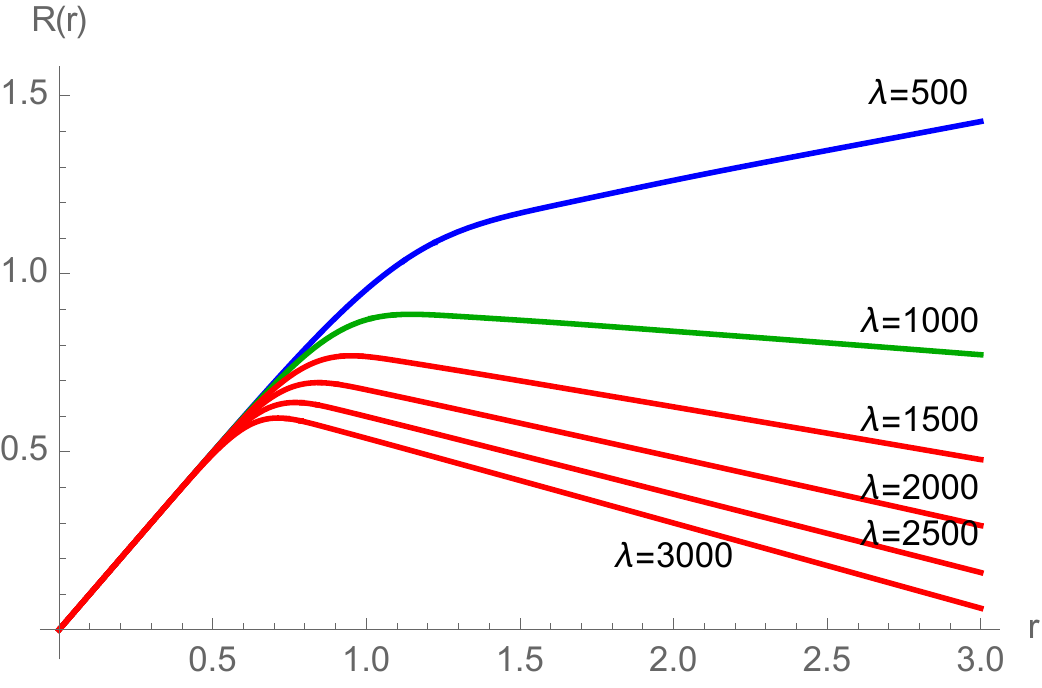}
	\caption{Examples of scalar lump solutions with $v=0.15$ and $\lambda$ increasing in steps of $500$ up to $\lambda=3000$. In all cases the scalar lump is confined within a horizon where $f=0.$ For the solution in blue ($\lambda=500$) the scalar field undergoes damped oscillations outside the horizon, where the universe asymptotes to de Sitter spacetime. For the green solution ($\lambda=1000$), the size of the 2-spheres $R(r)$ has a maximum outside of the horizon, and then shrinks again to zero, where a big crunch curvature singularity resides. In this case the lump is surrounded outside of its horizon by a collapsing universe. For the red solutions ($\lambda \geq 1500$) the turnaround in $R(r)$ already occurs within the horizon, and in these cases the scalar lump is surrounded by a black hole.}
	\label{fig:v150profiles}
\end{figure}

We find solutions of three types: those for which $R(r)$ increases monotonically, those for which $R(r)$ has a maximum and turns around beyond the horizon, and finally those for which $R(r)$ turns around already before a horizon is reached. Examples of solutions with a single interpolation ($n=1$) are provided in Fig. \ref{fig:v150profiles}. In this figure, the potential minima are fixed at $v=\pm 0.15$ while $\lambda$ takes values between $\lambda=500$ and $\lambda=3000$ in steps of $500.$ When $\lambda=500$ the solution has a monotonically increasing sphere radius $R$ -- we show such solutions in blue. The scalar field starts a little above the potential minimum at $\varphi=-0.15,$ moves across the potential barrier at $\varphi=0$ and then undergoes damped oscillations around the second potential minimum at $\varphi=+0.15.$ Meanwhile $f(r)$ decreases from the regular origin where $f(0)=1,$ passes through zero (thus indicating the location of the cosmological horizon) before tending to minus infinity according to $f(r) \to -\frac{V(0.15)}{3}r^2.$ This solution describes a scalar lump concentrated within the cosmological horizon. Asymptotically the solution is de Sitter spacetime with the scalar field settling down at a potential minimum. The spatial 2-spheres keep growing without bound.

As $\lambda$ is increased to $1000$ the character of the solution changes, as now $R$ is no longer monotonic\footnote{In Schwarzschild coordinates this obstruction would manifest itself by a divergence in the metric function~$\delta(r).$}. At first the solution evolves in much the same way as for the previous case: the scalar field starts off a little above the potential minimum at $\varphi=-0.15$ and moves across the barrier before undergoing damped oscillations around the second minimum at $\varphi=+0.15.$ Meanwhile $f(r)$ passes through zero at around $r_h\approx 1.08.$ But at $r\approx 1.14$ the size of the 2-spheres $R$ reaches a maximum before turning around and decreasing towards smaller values again. Such solutions, where $R$ turns around outside of the horizon, are shown in green in our graphs. As discussed in the previous section, $R$ is not allowed to have a minimum. Hence, once it decreases it will inevitably reach zero (at $r=r_s$), which is the location of a curvature singularity.  Thus behind the cosmological horizon, this solution corresponds to a collapsing universe, despite the presence of positive vacuum energy -- it is the spatial curvature of the 2-sphere that causes the universe to collapse. Near the singularity the solution becomes approximately of Kasner type, with anisotropic contraction of the spatial part of the metric and a logarithmic divergence of the scalar field. More precisely, the solution tends to Kasner form with a parameter $b,$
\begin{align}
& \varphi(r) \to \frac{3\varphi_1}{3b+4}\ln (r_s-r)\,, \quad f(r) \to -(r_s-r)^{\frac{6b+2}{3b+4}}\,, \quad R(r)\to (r_s-r)^{\frac{-\frac{3}{2}b+1}{3b+4}}\,, \nonumber \\ & b^2+\frac{2}{3}\varphi_1^2 = \frac{4}{9}\,,
\end{align}
where $\varphi_1$ is a constant and $r,$ being behind the horizon, is more aptly thought of as a time coordinate. However, this collapsing universe is shielded from the scalar lump by the horizon, and thus remains invisible to an observer at the origin.

Once $\lambda$ is further increased, the solution changes its character once more: at $\lambda=1500$ we can see that $R$ already turns around inside of the horizon. More precisely, $R$ turns around at $r\approx 0.94$ while the horizon is located at $r_h\approx 0.98.$ Such solutions, where $R$ turns around before a horizon is reached, are shown in red in our graphs. In this case, the horizon is more properly identified as a black hole horizon, and the solution is to good accuracy that of Schwarzschild-de Sitter spacetime, though with a non-trivial scalar field included. The scalar field once again diverges to infinity in Kasner fashion as the singularity at $R=0$ is approached. Solutions of this type thus describe a scalar lump surrounded in all directions by a black hole, a rather claustrophobic situation!

As $\lambda$ is increased further, the singularity at $R=0$ moves closer and closer to the horizon $f=0,$ without however reaching the horizon, i.e. the singularity remains shielded by a horizon. Thus no further types of solutions (with a single interpolation of the scalar field) exist within the present ansatz.

\begin{figure}[h]
	\centering
	\includegraphics[width=0.6\textwidth]{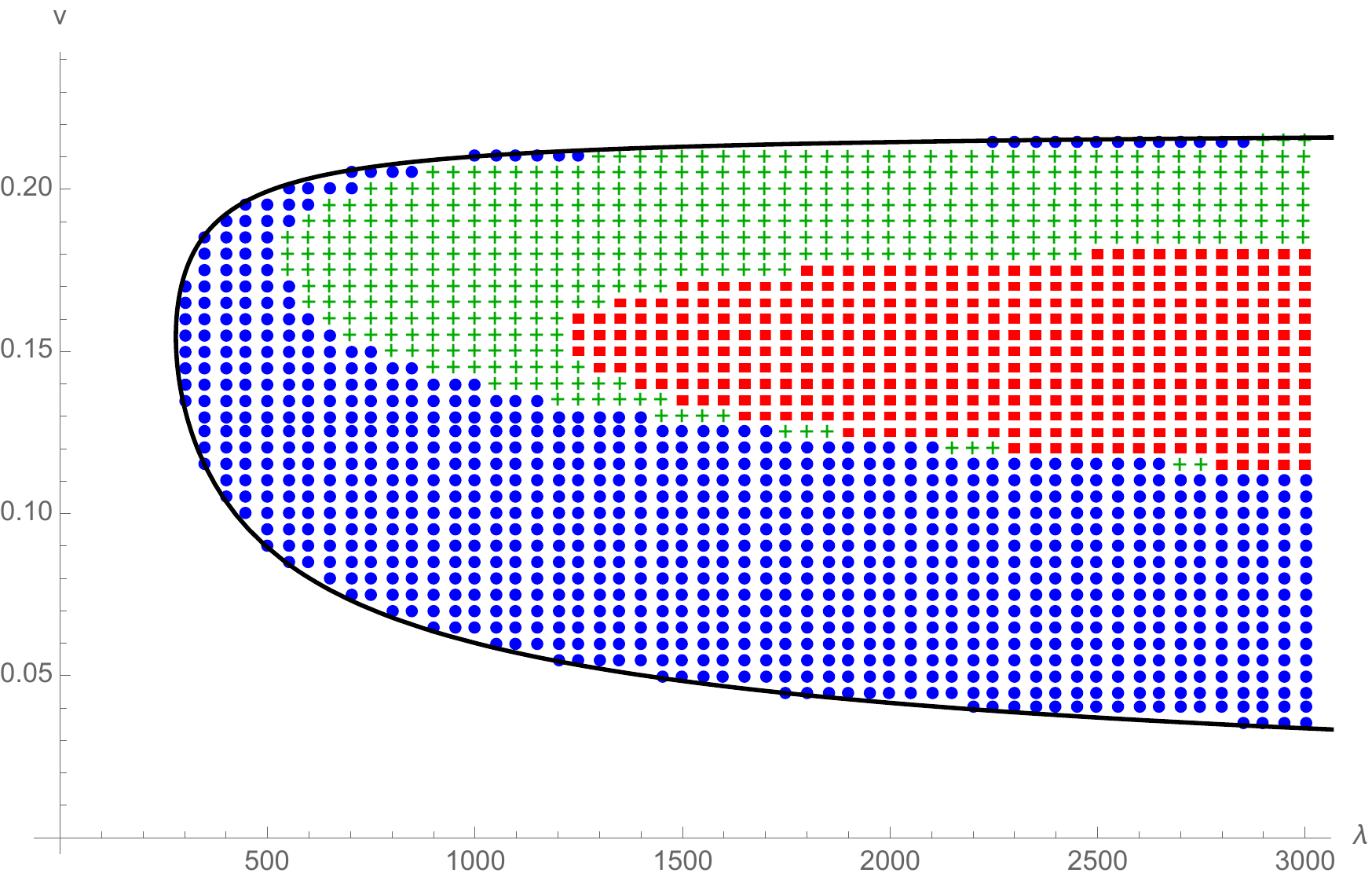}
	\caption{This summary graph shows all of the solutions we have optimised for which there is a single interpolation of the scalar field across the potential barrier ($n=1$). As one can see, the nature of the solutions depends both on the re-scaled coupling constant $\lambda$ and on the location $v$ of the vacua in the potential. Blue dots indicate the presence of a solution with monotonic $R(r),$ green plusses those in which $R(r)$ has a maximum outside of the horizon and red rectangles those with a maximum of $R(r)$ inside the horizon. To the left of the black line \eqref{lambdacritical} no interpolating solutions exist, while to the right of this line we expect solutions to exist up to arbitrarily large values of $\lambda.$}
	\label{fig:n1}
\end{figure}

In order to obtain a global overview of the existence of solutions with $n=1$, we have performed a survey by optimising solutions in a grid of $\lambda-v$ values. The result is shown in Fig. \ref{fig:n1}. The black line corresponds to the critical values specified by the analytic formula \eqref{lambdacritical} with $n=1,$ to the left of which no interpolating solutions exist. Solutions that are very close to this line exhibit only a small interpolation of the scalar field across the top of the potential barrier. As $\lambda$ is increased, we reach the blue-green transition at which the solutions asymptote to the Nariai solution with constant $R$ at large $r.$ The solutions in green have a turnaround in $R$ beyond the horizon, and thus develop a singularity interpreted as the big crunch of a collapsing universe. The limiting case corresponds to the situation where $R$ turns around precisely at the horizon, which corresponds to the green-red boundary. Beyond these values of $\lambda$ the solutions correspond to the ``claustrophobic'' scalar lumps surrounded by a black hole.

\begin{figure}[h]
	\centering
	\includegraphics[width=0.24\textwidth]{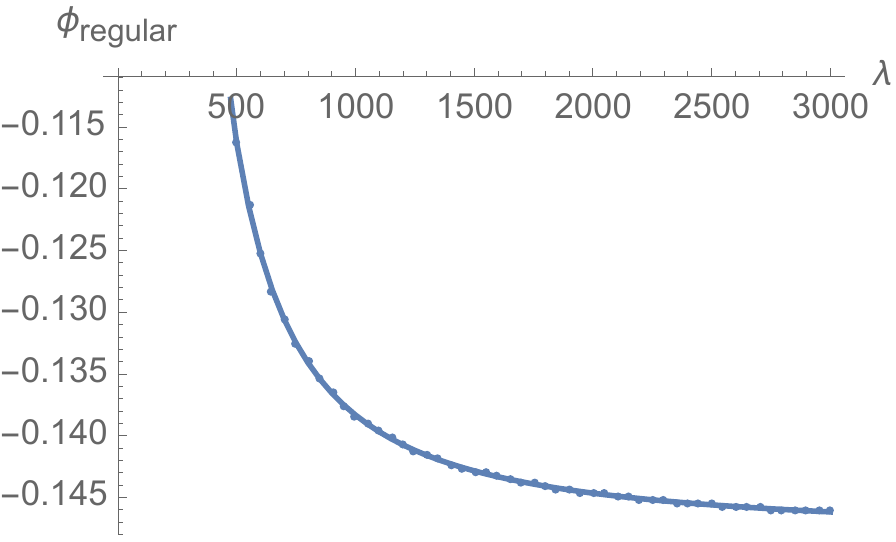}
	\includegraphics[width=0.24\textwidth]{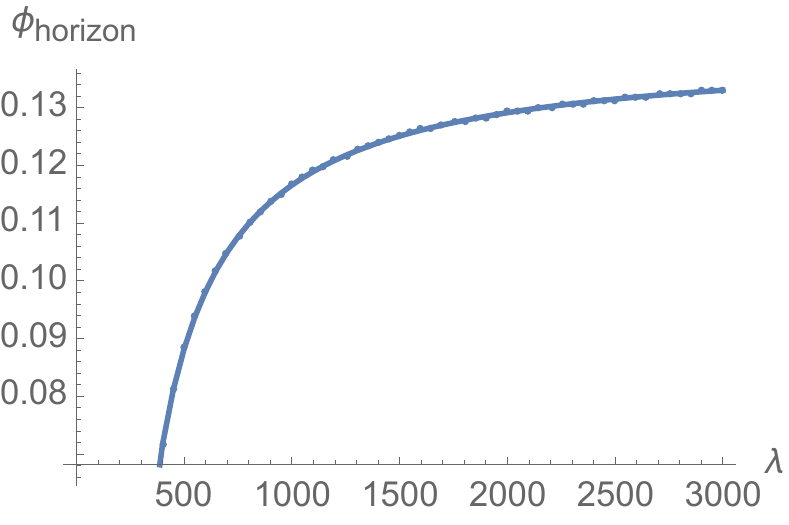}
	\includegraphics[width=0.24\textwidth]{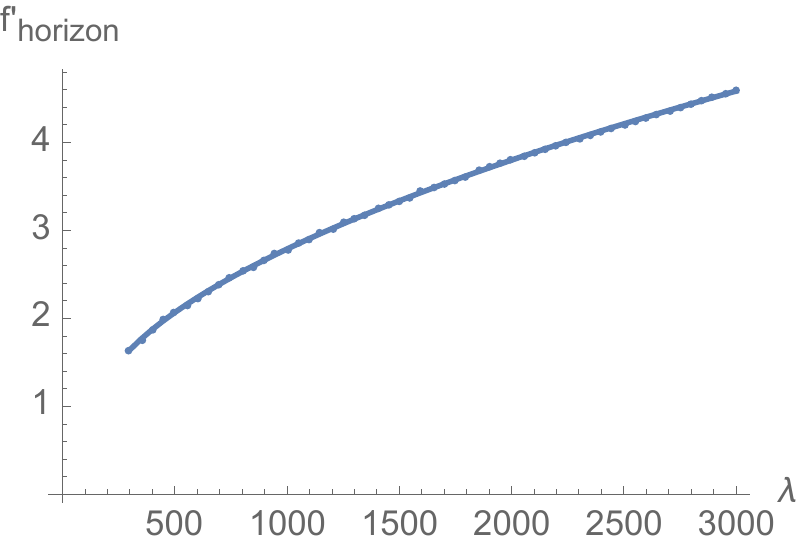}
        \includegraphics[width=0.24\textwidth]{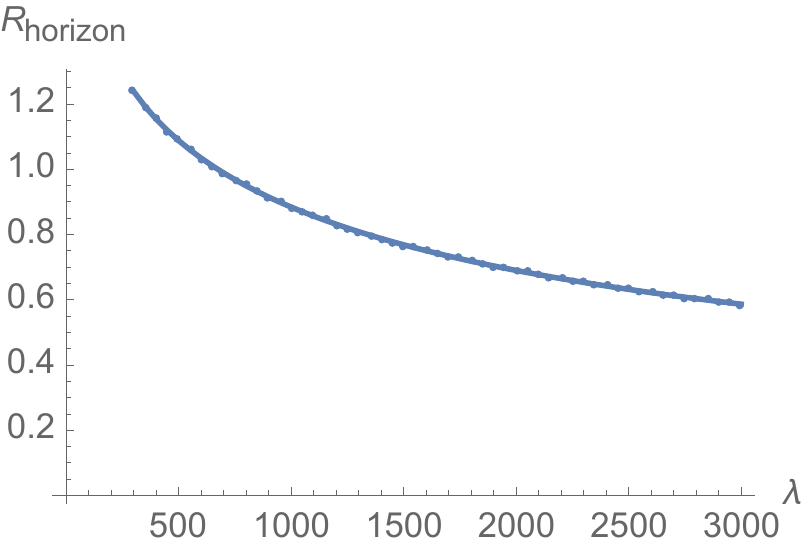}
	\caption{Graphs of the optimised parameters at the regular origin $\varphi_0=\varphi_{regular}$ and at the horizon, for fixed $v=0.15$ and a range of values of $\lambda$. }
	\label{fig:v150}
\end{figure}

\begin{figure}[h]
	\centering
	\includegraphics[width=0.24\textwidth]{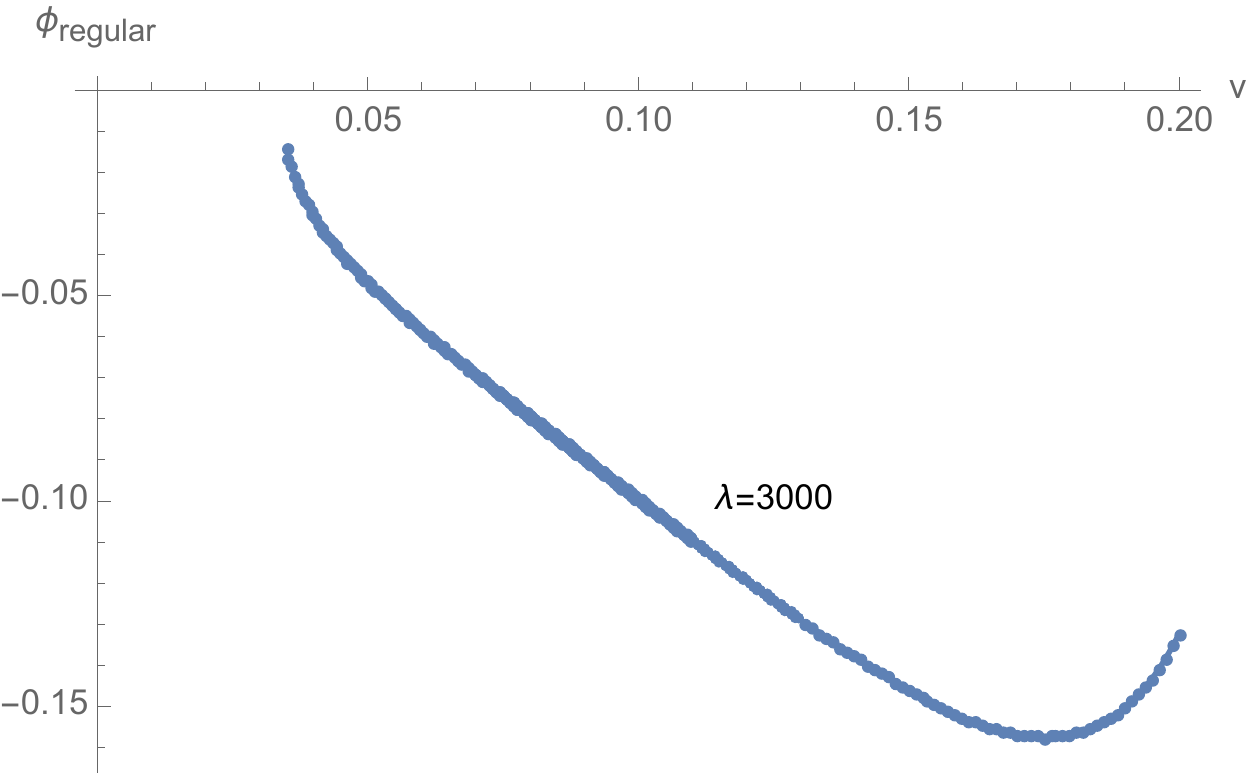}
	\includegraphics[width=0.24\textwidth]{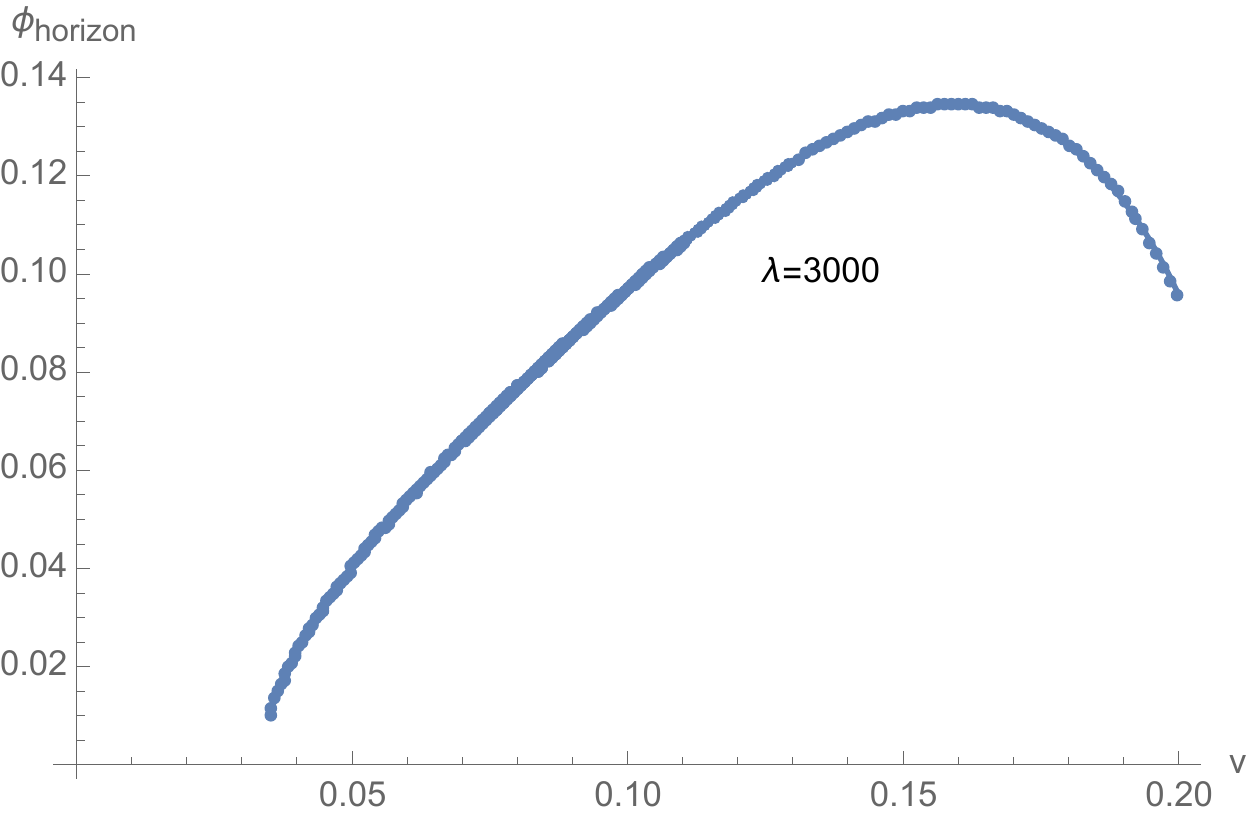}
	\includegraphics[width=0.24\textwidth]{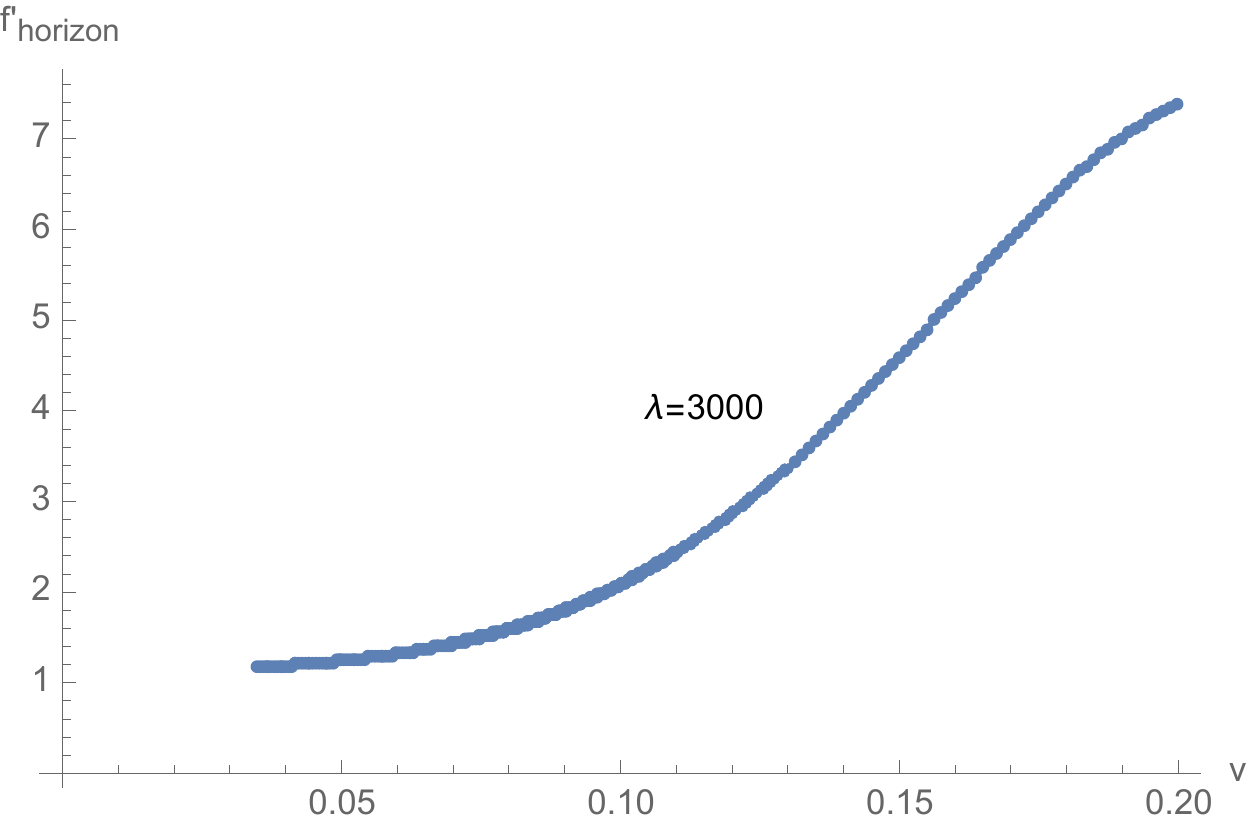}
        \includegraphics[width=0.24\textwidth]{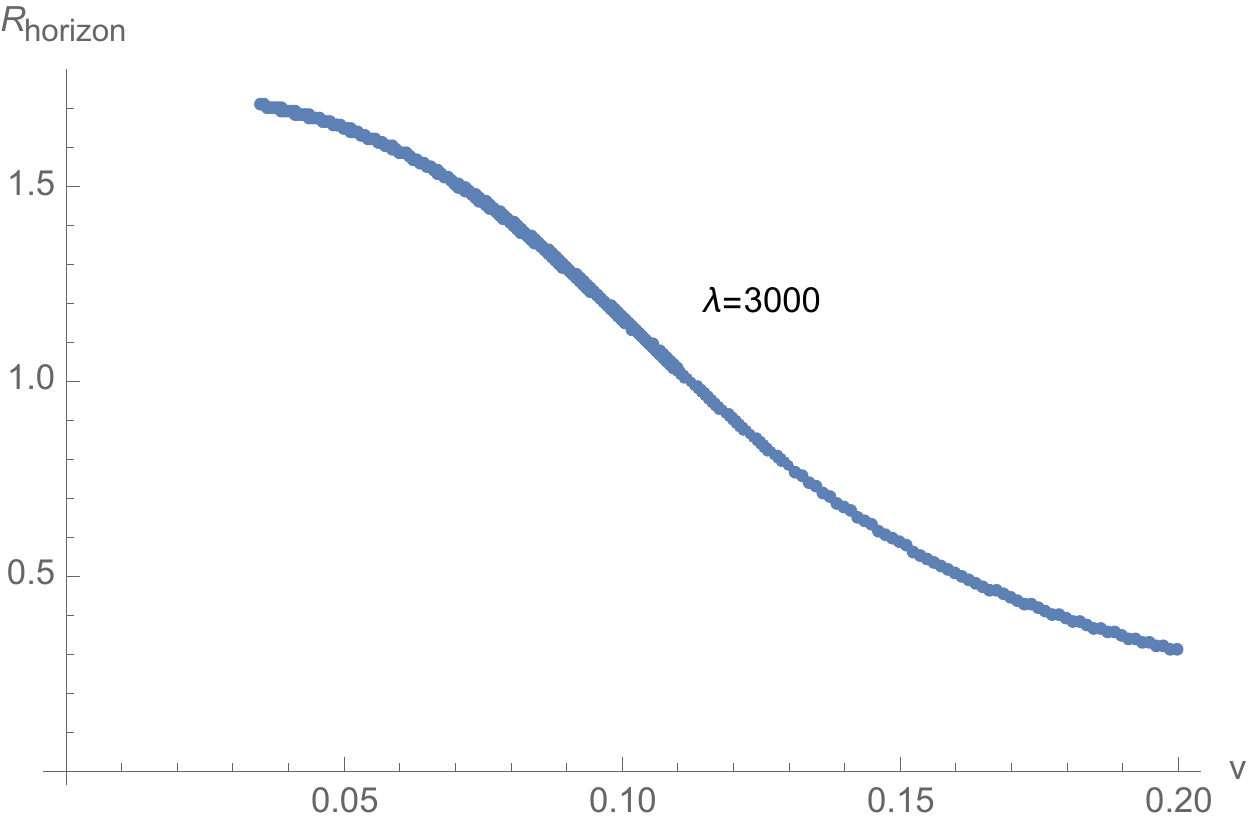}
	\caption{Graphs of the optimised parameters at the regular origin $\varphi_0=\varphi_{regular}$ and at the horizon, this time for fixed $\lambda=3000$ and a range of values of $v$. }
	\label{fig:lambda3000}
\end{figure}

In Figs. \ref{fig:v150} and \ref{fig:lambda3000} we plot the optimised values of the scalar field $\varphi_0$ at the regular origin, together with the optimised parameters at the horizon $\varphi_h, f_h', R_h$ for slices through the data, more precisely for a slice at constant $v=0.15$ in Fig. \ref{fig:v150} and at constant $\lambda=3000$ in Fig. \ref{fig:lambda3000}. For fixed $v$ and increasing $\lambda$ it is clear from our preceding discussion that the evolution of the solutions is monotonic, as the graphs readily verify. However, at constant $\lambda$ and changing $v$ we can see that the optimised parameters do not evolve monotonically.

\begin{figure}[h]
	\centering
	\includegraphics[width=0.3\textwidth]{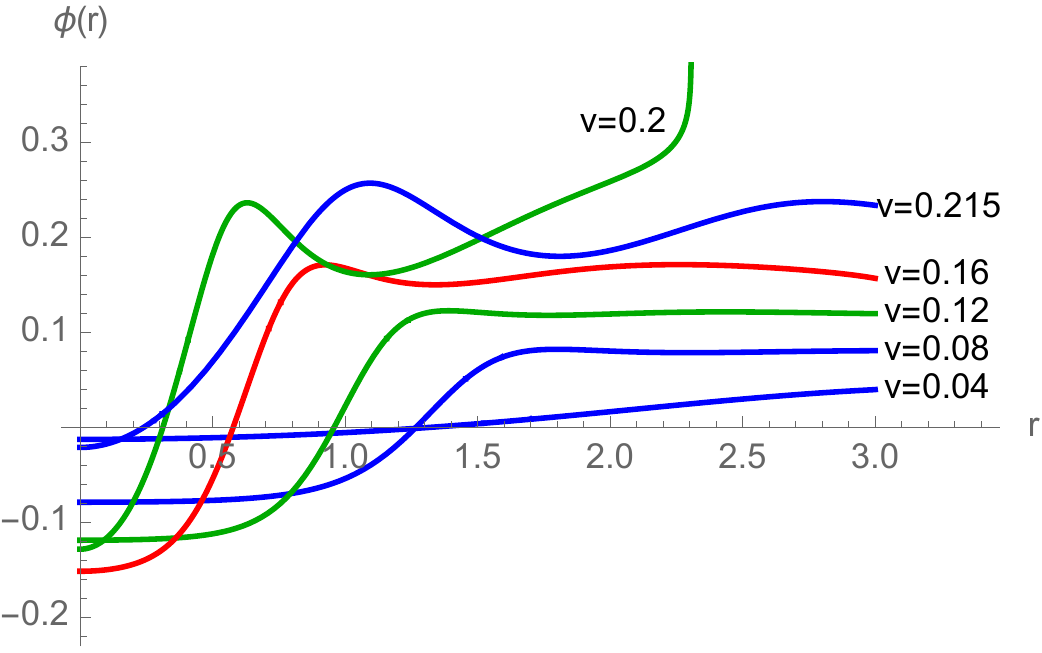}
	\includegraphics[width=0.3\textwidth]{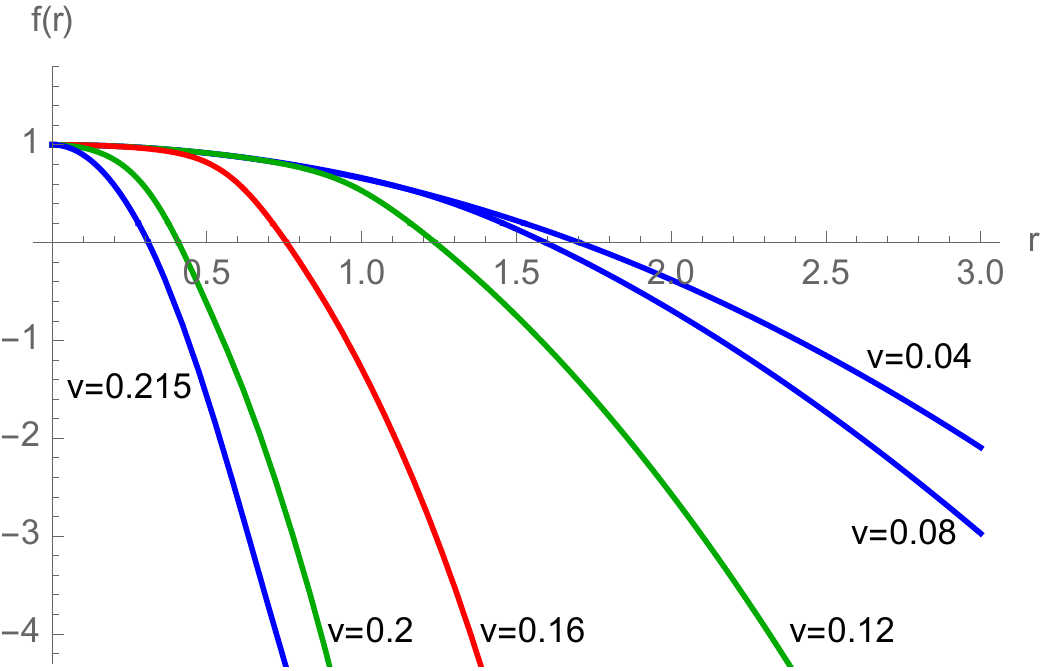}
	\includegraphics[width=0.3\textwidth]{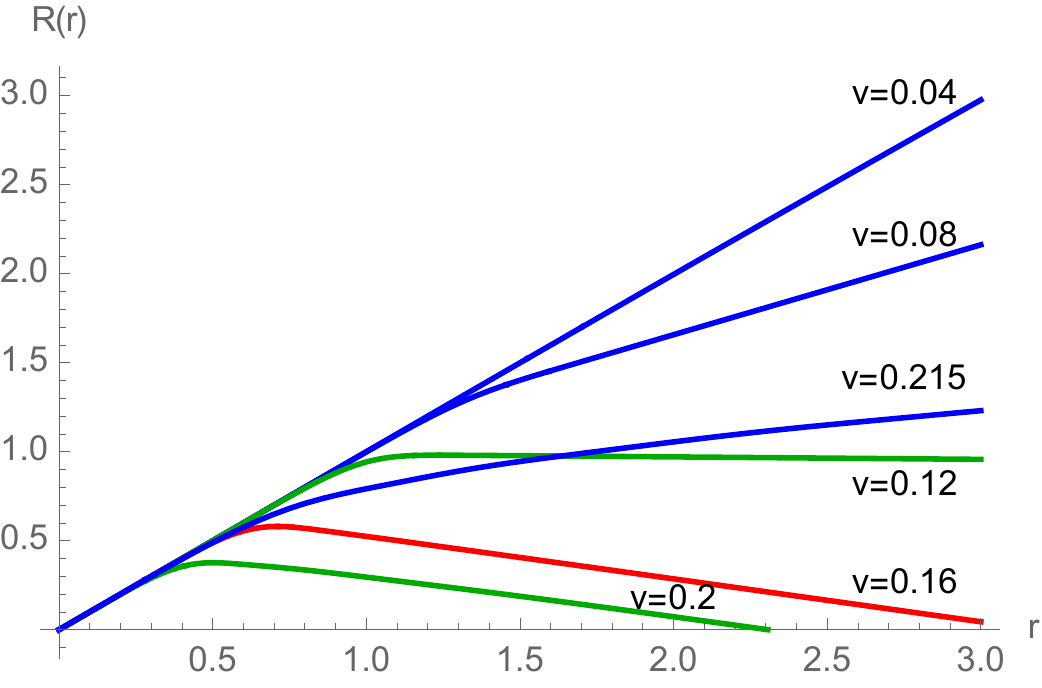}
	\caption{Examples of solutions with fixed $\lambda=2250$ and the indicated values of $v$. These examples provide a vertical slice through the summary graph in Fig. \ref{fig:n1}.}
	\label{fig:lambda2250profiles}
\end{figure}

Explicit examples of solutions with constant $\lambda$ and varying $v$ are shown for illustration in Fig. \ref{fig:lambda2250profiles}. When analysing these graphs one should bear in mind that the potential minima are located at $\varphi = \pm v,$ and thus for different values of $v$ it is quite natural that the starting and end values of $\varphi$ change significantly. Note that for the solution with $v=0.2$ one can already see the Kasner-like divergence of the scalar field in the approach to the spacetime singularity $R=0$ at $r_s\approx 2.3$.

\begin{figure}[h]
	\centering
	\includegraphics[width=0.6\textwidth]{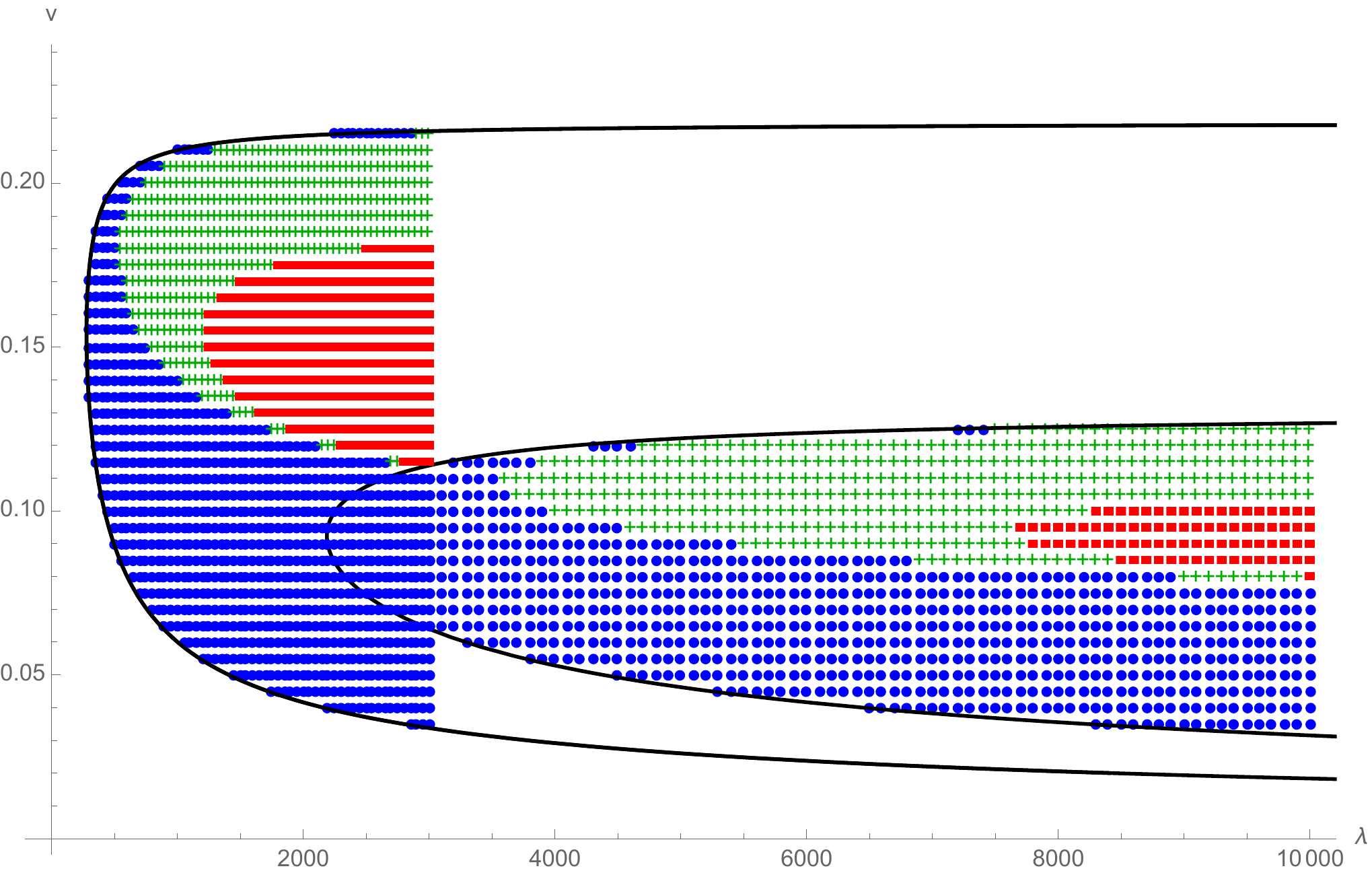}
	\caption{This summary graph shows all of the solutions we have optimised for which there are two interpolations of the scalar field across the potential barrier ($n=2$). Once again, the nature of the solutions depends both on the re-scaled coupling constant $\lambda$ and on the location $v$ of the vacua in the potential. Blue dots indicate the presence of a solution with monotonic $R(r),$ green plusses those in which $R(r)$ has a maximum outside of the horizon and red rectangles those with a maximum of $R(r)$ inside the horizon. For convenience we have superimposed the solutions with $n=1$ that were already shown in Fig. \ref{fig:n1}. We expect solutions to keep existing arbitrarily far to the right of the critical black lines, drawn here for both $n=1$ and $n=2.$}
	\label{fig:n2}
\end{figure}

Analogous results may be obtained for solutions that interpolate twice across the potential barrier, i.e. for the case that $n=2.$ As is evident from the analytic formula \eqref{lambdacritical} such solutions can only exist at larger values of $\lambda$ and for a more restricted range of $v.$ We have performed a similar survey of solutions as in Fig. \ref{fig:n1}, and this is shown in Fig. \ref{fig:n2}. Note that in this figure we have also included the optimised solutions with a single interpolation, for comparison. We find a very similar pattern of solution when $n=2,$ i.e. once again we find solutions with monotonic $R$ (in blue), with a turnaround in $R$ beyond the horizon (in green) and a turnaround in $R$ within the horizon (in red).

\begin{figure}[h]
	\centering
	\includegraphics[width=0.31\textwidth]{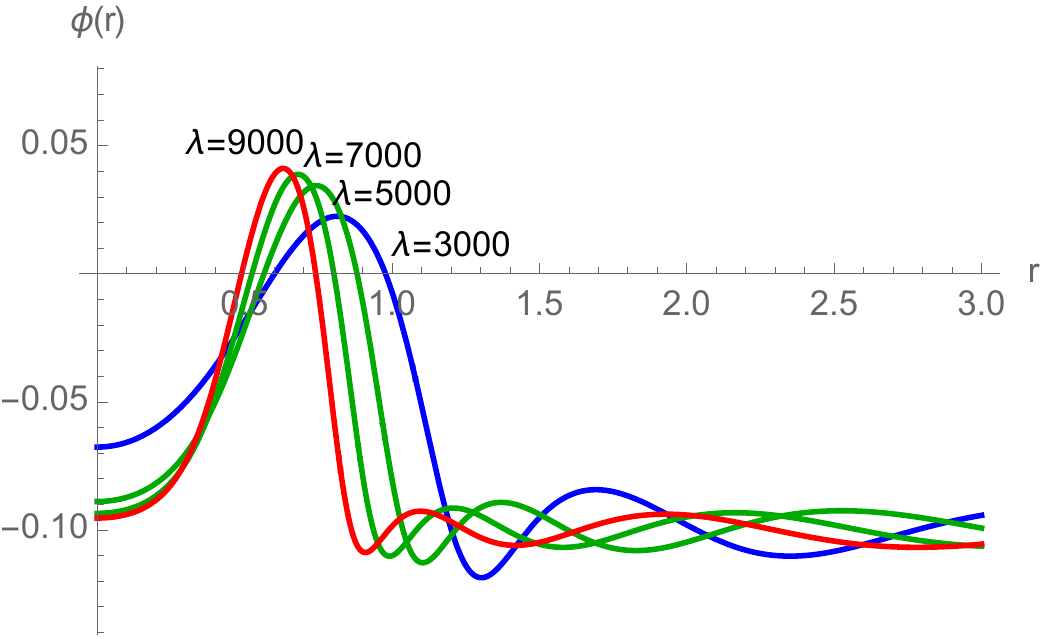}
	\includegraphics[width=0.31\textwidth]{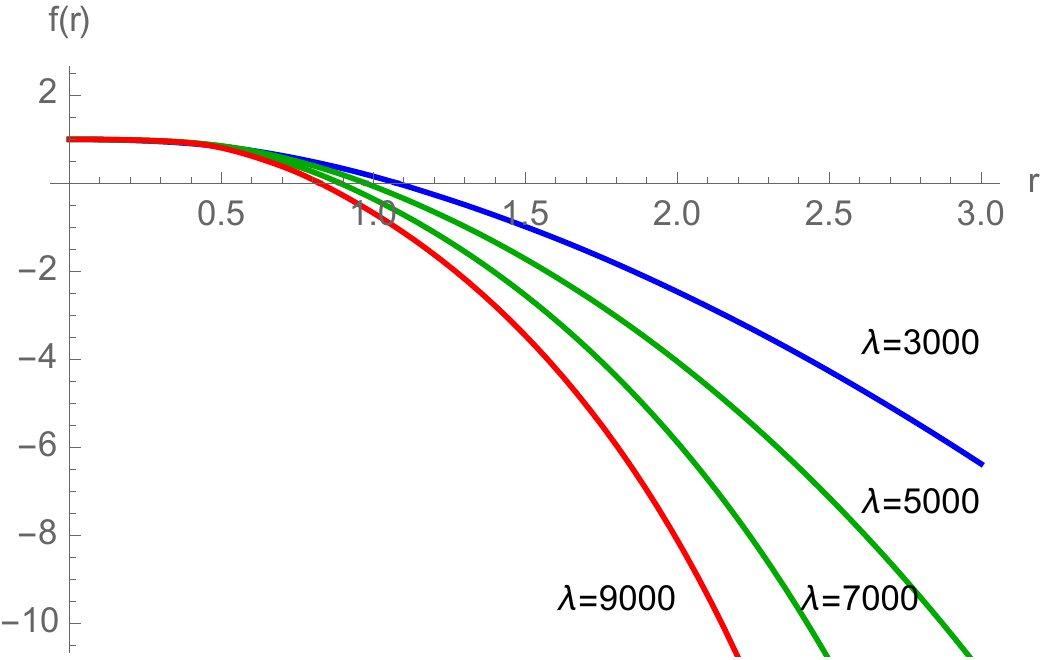}
	\includegraphics[width=0.31\textwidth]{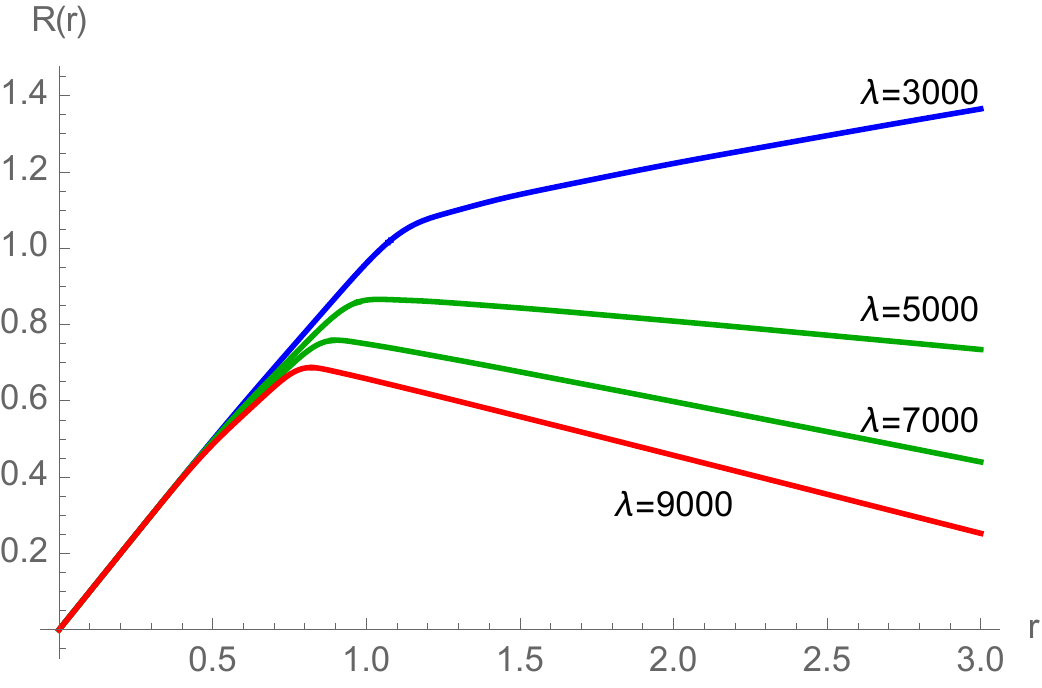}
	\caption{Examples of scalar lump solutions with two interpolations of the scalar field ($n=2$) for fixed $v=0.1$ and the indicated values of $\lambda$. Except for the fact that the scalar field moves across the potential barrier and back again, the behaviour of these solutions is analogous to the case where $n=1,$ shown above in Fig. \ref{fig:v150profiles}.}
	\label{fig:n2v100profiles}
\end{figure}

\begin{figure}[h]
	\centering
	\includegraphics[width=0.24\textwidth]{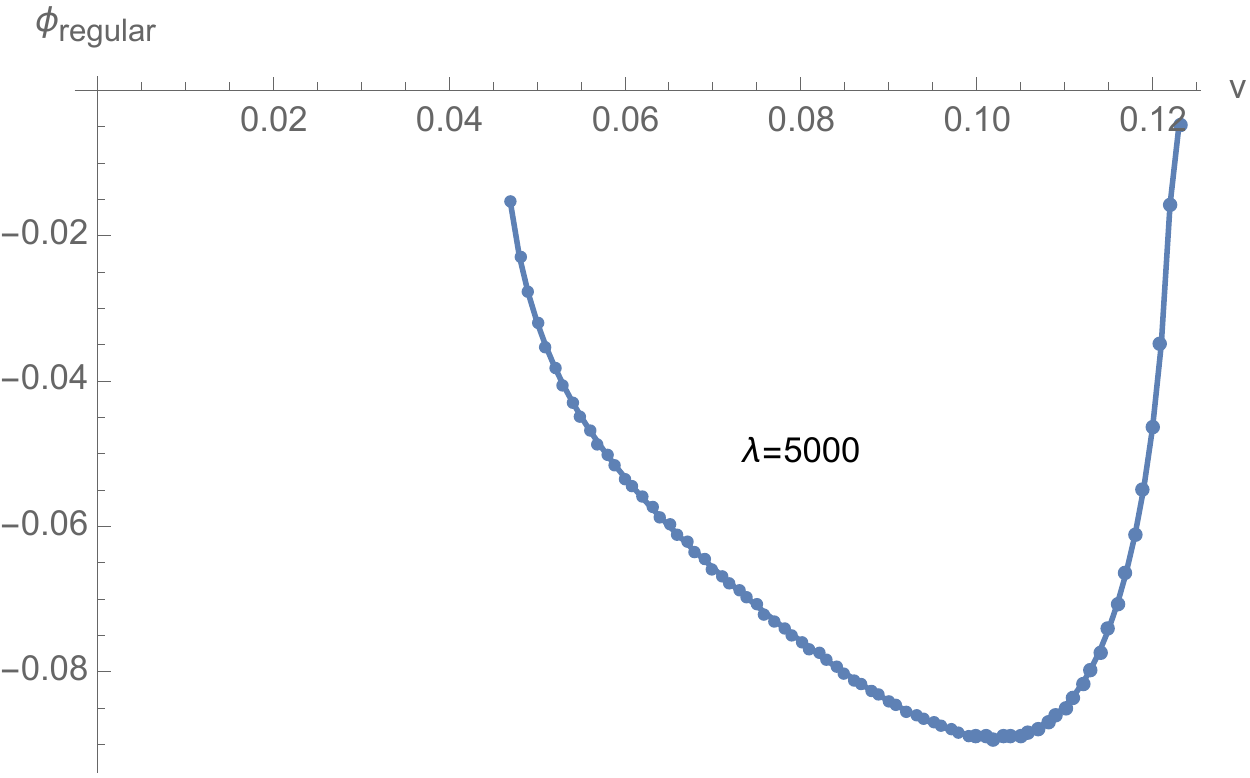}
	\includegraphics[width=0.24\textwidth]{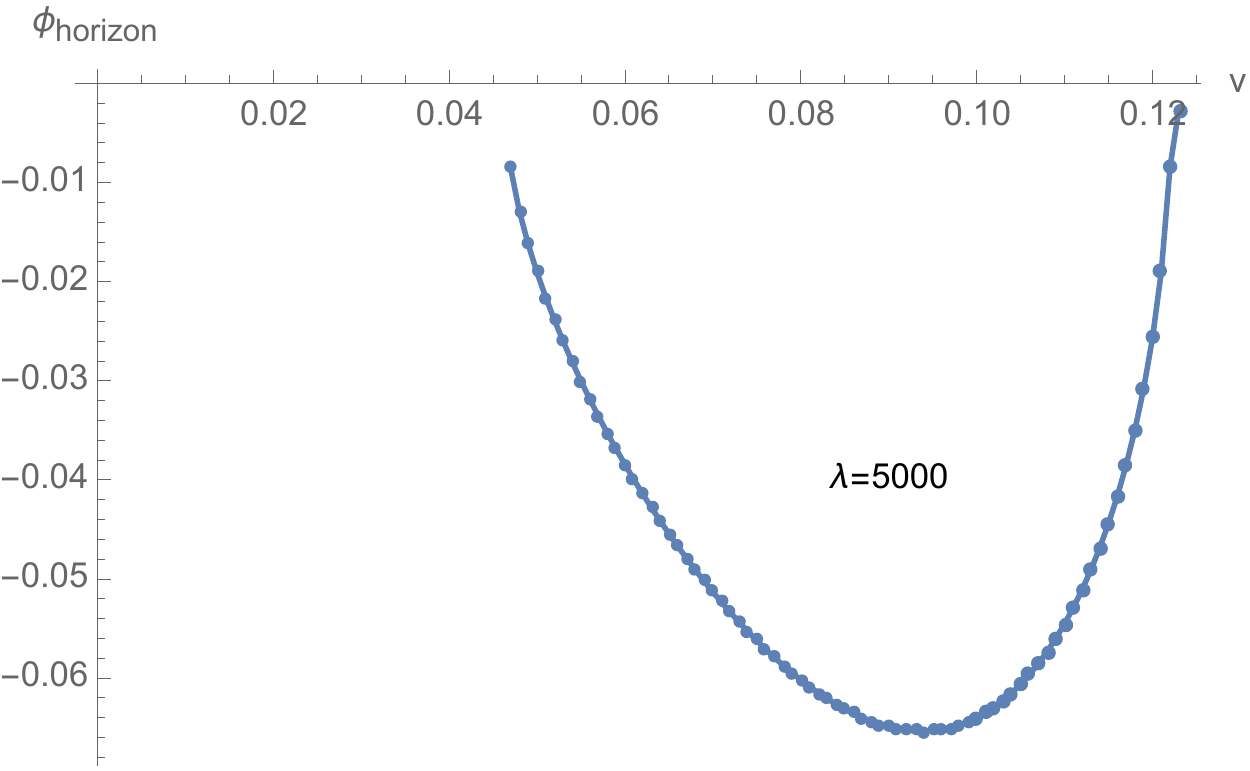}
	\includegraphics[width=0.24\textwidth]{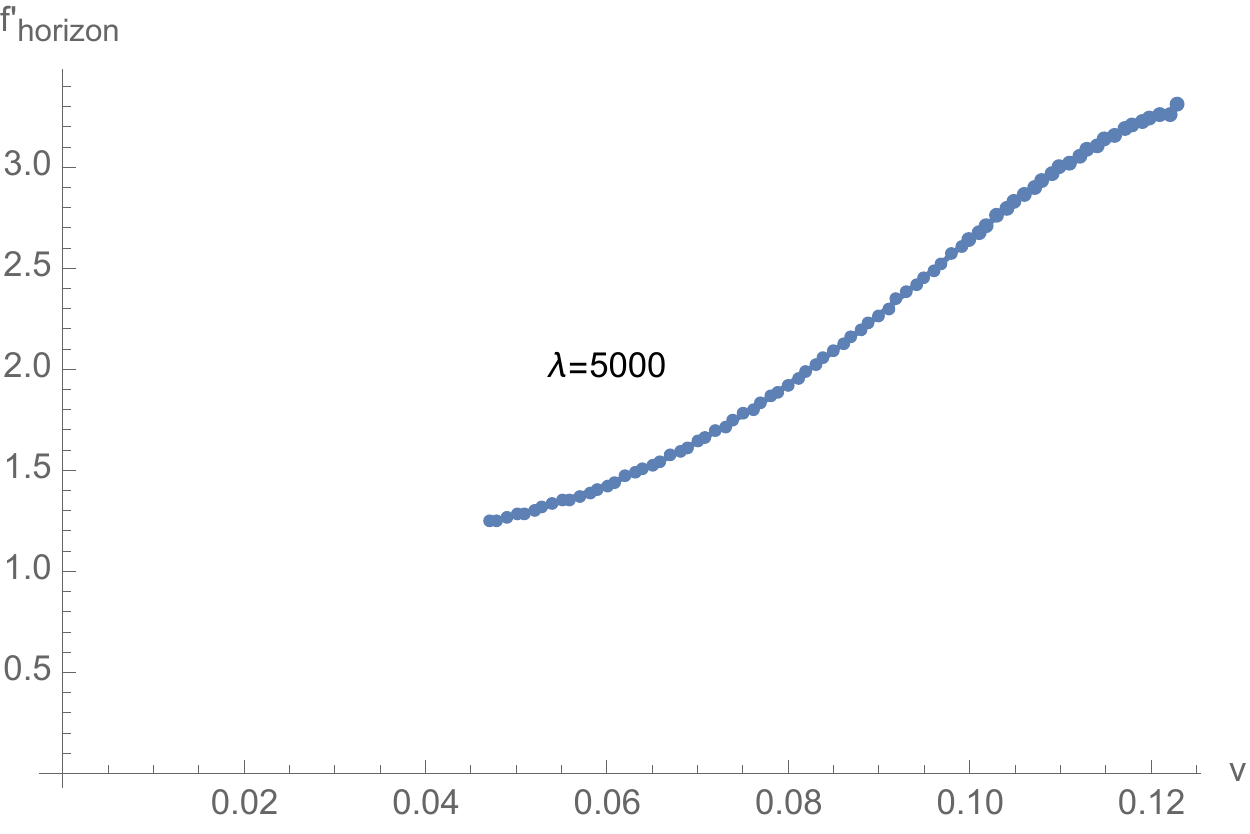}
        \includegraphics[width=0.24\textwidth]{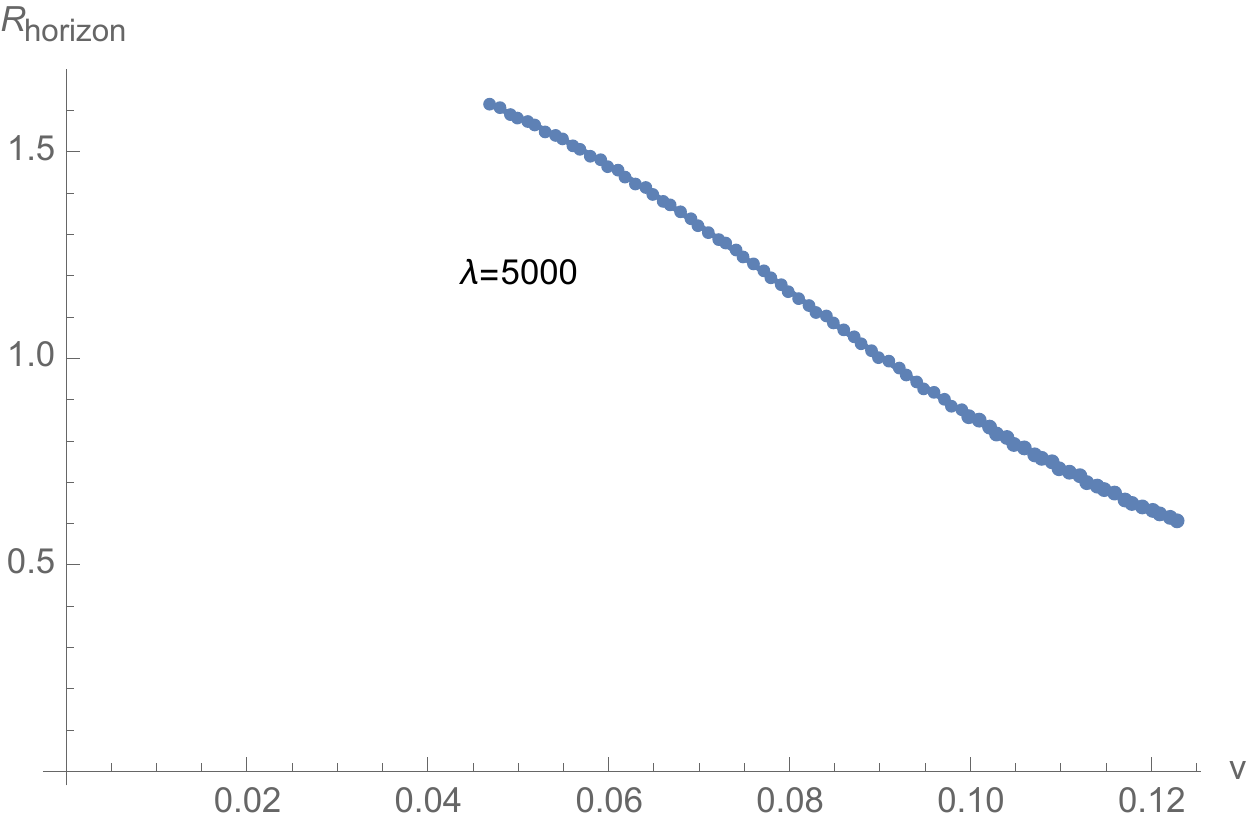}
	\caption{Graphs of the optimised parameters at the regular origin $\varphi_0=\varphi_{regular}$ and at the horizon, for solutions with two interpolations $n=2,$ for fixed $\lambda=5000$ and a range of values of $v$.  }
	\label{fig:n2lambda5000}
\end{figure}

Several explicit examples with $n=2$ are shown in Fig. \ref{fig:n2v100profiles}. In these examples $v=0.10$ is held constant and $\lambda$ is increased in steps of $2000$ from $\lambda=3000$ to $\lambda=9000.$ Here one can explicitly see the monotonic progression of the solutions as the coupling constant is increased. For reference, we also plot the optimised parameter values at the regular origin and at the horizon for a slice through the data at constant $\lambda=5000$ and varying $v,$ see Fig. \ref{fig:n2lambda5000}. In this case, we have solutions with monotonic $R$ at both ends of the allowed $v$ range, while as we turn towards the more central $v$ values we first reach solutions with a $R$ turnaround outside of the horizon and eventually within the horizon, cf. the location of these solutions in Fig. \ref{fig:n2}. One can see that near the ends of the allowed $v$ range the solutions exhibit only a very small interpolation of the scalar field across the potential barrier and back, while towards the middle of the allowed range in $v$ the solutions start and end rather close to the potential minima. This implies that for such solutions to exist it does not actually matter whether or not the potential has minima, the important feature is the potential barrier. This is reminiscent of Coleman-de Lucchia instantons, which also start and end away from potential minima and require a barrier for their existence. Note however that our solutions here are more accurately described as solitons within the horizon (and collapsing universes outside).

We fully expect the observed patterns for $n=1$ and $n=2,$ i.e. nested regions in which $R$ turns around closer and closer to the origin as the coupling constant $\lambda$ is increased, to continue for all higher interpolating solutions with $n>2.$

%%%%%%%%%%%%%%%%%%%%%%%%%%%%%%%%%%%%%%%%%%%%%%%%%

%%%%%%%%%%%%%%%%%%%%%%%%%%%%%%%%%%%%%%%%%%%%%%%%%

\section{Causal structure} \label{sec:causal}

Geometric properties of space-times with Lorentzian signature are strongly related to how two separated
points communicate with each other, which is centralised under the notion of causal structure. We can classify points into regions of which the most prominent are normal (or untrapped), trapped, and
marginally trapped, where the last are essential for
the definition of horizons. Since in this article we study solutions that typically do not allow for a notion of asymptotic flatness, we must define those regions locally. Hence we adopt the formalism of dynamical
\cite{Ashtekar:2003} or trapping horizons \cite{Hayward:1994} that become isolated horizons
in static cases and, therefore, agree with the location of event
horizons in simple space-times such as spherically symmetric ones \cite{Ashtekar:2000}.
 This quasi-local description only
presumes time-orientability and orientability as global properties.

The classification is carried out using the expansion parameter $\theta$ along two
future-directed null-congruences, $l$ and $n$, which span the lightcone locally.
Due to the symmetries of the metric \eqref{metricansatz}, there exist two spheres of light along the null congruences, which are
radial null geodesics,
that pass through every point on the manifold, either called ingoing or outgoing.
Following \cite{Hayward:1994},
we define the dual-null foliation according to $l$ and $n$ as an
embedding $\mathcal{M}\to\mathcal{X}_+\times\mathcal{X}_-\times\mathcal{S}$,
where the half-open intervals
$\mathcal{X}_-=[0,x_-^0)$ and $\mathcal{X}_+=[0,x_+^0)$ span the evolution space along the
null-congruences with $\mathcal{S}$ being a spatial, two-dimensional submanifold with induced metric $h$.
The associated coordinates $x_-=t-r$ and $x_+=t+r$ will refer to in- and outgoing light-cone coordinates.
From the above embedding it is clear that the area of ingoing spheres of light decreases, while that of
outgoing spheres of light increases.
This behaviour is exactly described by the expansion parameter $\theta^W(p)$ at a point $p$ along
a vector field $W$. More precisely, $\theta^W(p)$ describes the change of an infinitesimal area $\delta A^W$
that is Lie-dragged along a vector field $W$ with affine parameter $s$ and induced spatial metric
$h$ on $\mathcal{S}$:
\begin{equation}\label{theta}
\theta^W(p)=\frac{1}{\delta A^W}\frac{\mbox{d}}{\mbox{d}s}\delta A^{W}=
h^{-1}(\mathcal{L}_Wh)=h^{ab}W^c\nabla_ch_{ab}.
\end{equation}
In the following, we consider $\theta^n(p)$ and $\theta^l(p)$ to chart the space-time's causal structure
similar to \cite{Liberati:2020}.
By forming the geometrically
invariant product $\theta^n\theta^l(p)$ at $p\in\mathcal{M}$
we can classify normal surfaces ($\theta^n\theta^l<0$) where
a reference area shrinks (i.e. $\theta<0$) along ingoing and expands (i.e. $\theta>0$)
along outgoing null-congruences, trapped surfaces
($\theta^n\theta^l>0$) where either both congruences are effectively ingoing (future trapped) or outgoing
(past trapped), and the marginally trapped region ($\theta^n\theta^l=0$ with one $\theta\neq0$); the case
where both $\theta\equiv0$ refers to extremal surfaces that are either maximal or minimal.

%Normal and trapped regions are foliated by the corresponding surfaces, a three-surface $\mathcal{H}=\bigcup_{t\in\mathbb{R}}\mathcal{S}_t$ foliated by closed marginally trapped surfaces $\mathcal{S}_t$ along the time direction $t$ is called a trapping horizon and provides a local and dynamical description that can be used to classify horizons.

To understand the causal structure better, we need to elaborate on the extremal surface studying
the behaviour of the areal radius $R(r)$. There
exist two types of extremal surfaces: in static space-times, minimal surfaces are given by $R''(r)>0$
and maximal surfaces by $R''(r)<0$, also called equator, all evaluated at $r=r_E$ for which $R(r)$
develops a local extremum. While horizons link trapped and normal
regions, extremal surfaces connect either normal with normal or trapped with trapped regions
\cite{Senovilla:2007} leading to
a simultaneous sign change in both expansions. Intuitively, it can be thought of interchanging the notion of
ingoing and outgoing which translates in the trapped region to a change in the direction of the trapping, i.e.
from past to future or vice versa.

In spherically symmetric space-times, like \eqref{metricansatz}, the location where $\theta^n\theta^l=0$
becomes the condition $g^{rr}(r)=f(r)=0.$ The normal and the trapped region
can then be read off from the sign of $f(r)$. Diagonal metrics develop coordinate singularities at the horizon caused by the absence of non-degenerate, off-diagonal terms that compensate the zeroes in $f(r)$.
Changing to the retarded Eddington-Finkelstein-Bardeen form
\begin{equation}\label{EFB}
g=-f(r)\mbox{d}x_+^{2}+2\mbox{d}x_+\mbox{d}r+R^2(r)\mbox{d}^2\Omega,
\end{equation}
obviates this problem. Another advantage is that this coordinate chart can be defined at each point
of the extended manifold. We
find the future-directed
null-congruences through the relations $g^{-1}(l,l)=g^{-1}(n,n)=0$ and $g^{-1}(l,n)=-2$
to be $l^a=(\partial_{x_+})^a+\tfrac{f(r)}{2}(\partial_r)^a$ and $n^a=-2(\partial_r)^a,$ where
the partial derivatives are understood as local basis. By
using \eqref{theta}, we can construct the explicit form for $\theta^n(r)$ and $\theta^l(r)$ to be
\begin{eqnarray}
\theta^n(r)&=&-\frac{4}{R(r)}R'(r),\\
\theta^l(r)&=&\frac{f(r)}{R(r)}R'(r).
\end{eqnarray}
From these equations, we can immediately see that $\theta^l(r)=0$ is zero at the roots of $f(r)$ as well as
at the local maximum of $R(r)$. While $f(r)=0$ determines the location of the horizon $r_H$,
$R'(r)=0$ locates the extremal surface at $r_E$ where
both expansions vanish simultaneously \cite{Mars:2003}. A refinement of the horizon structure is achieved by
another geometrical invariant: $\mathcal{L}_Z\theta^W(p)$ involves the Lie derivative of the
expansion $\theta^W(p)$ along a vector field $Z$ evaluated at the marginally trapped surface. In our case:
\begin{eqnarray}\label{lietheta}
%\mathcal{L}_l\theta^n(r)&=&\frac{2}{R(r)}\left(\frac{R'(r)^2}{R(r)}-R''(r)\right),\\
\mathcal{L}_n\theta^l(r)&=&-\frac{2}{R(r)}\left(f'(r)R'(r)-\frac{f(r)R'(r)}{R(r)}+f(r)R''(r)\right).
\end{eqnarray}
Since the $\mathcal{L}_l\theta^n(r)$ can be shown to be insensitive on the slope of $f(r)$ at the horizon,
we will only consider $\mathcal{L}_n\theta^l(r)$.
The sign of $\mathcal{L}_n\theta^l(r)$ at $r_H$ determines the direction of trapping, either in outgoing direction
$\mathcal{L}_n\theta^l(r)>0$ (inner/cosmological horizon) or in ingoing direction
$\mathcal{L}_n\theta^l(r)<0$ (outer/black hole-like horizon), all evaluated at the horizon.
Note, only the first term in \eqref{lietheta}
contributes because the remaining terms are proportional to $f(r)$ which vanishes at the horizon.

\begin{figure}
  \centering
{\includegraphics[width=0.31\linewidth]{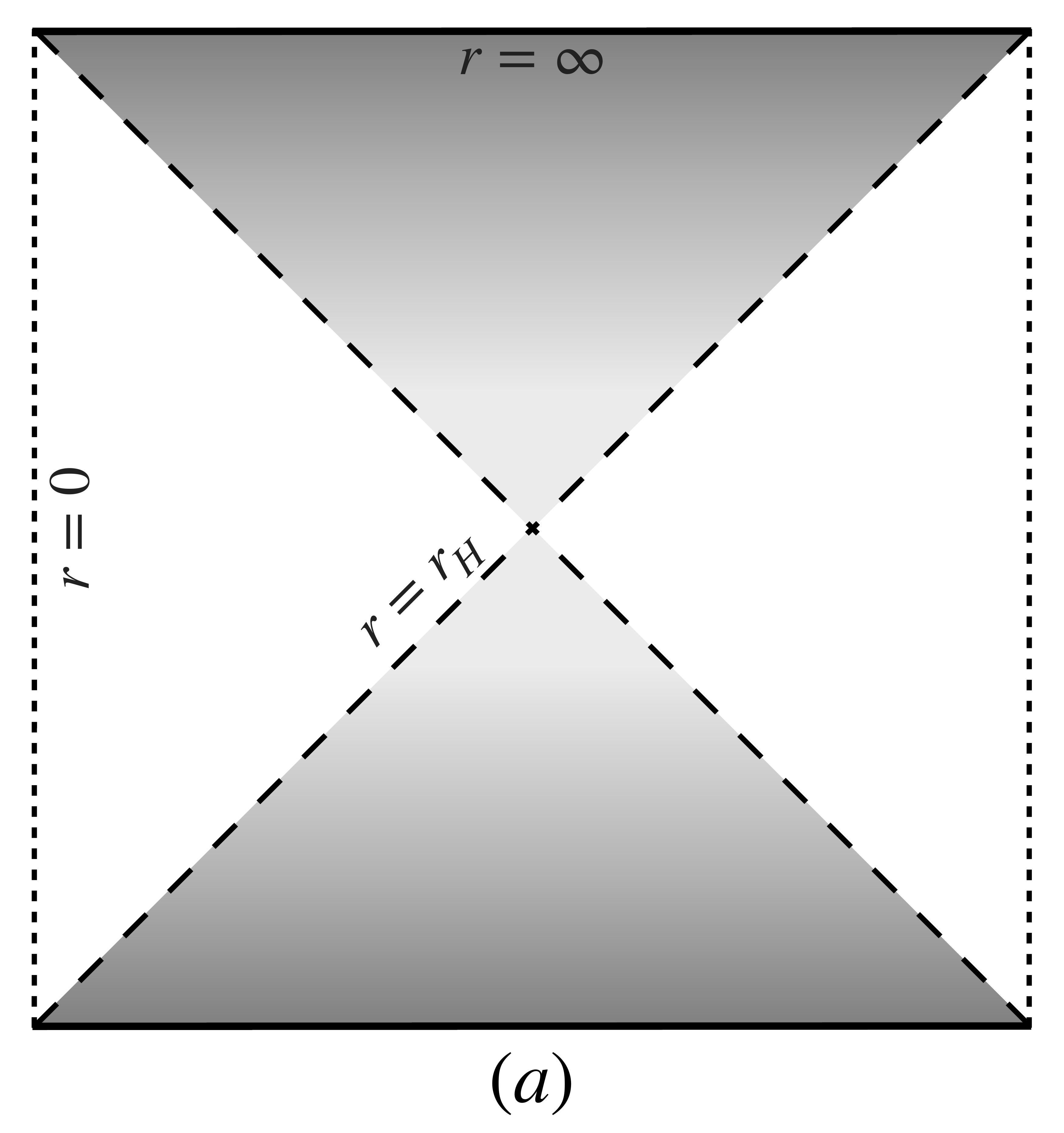}}%
  \quad
{\includegraphics[width=0.31\linewidth]{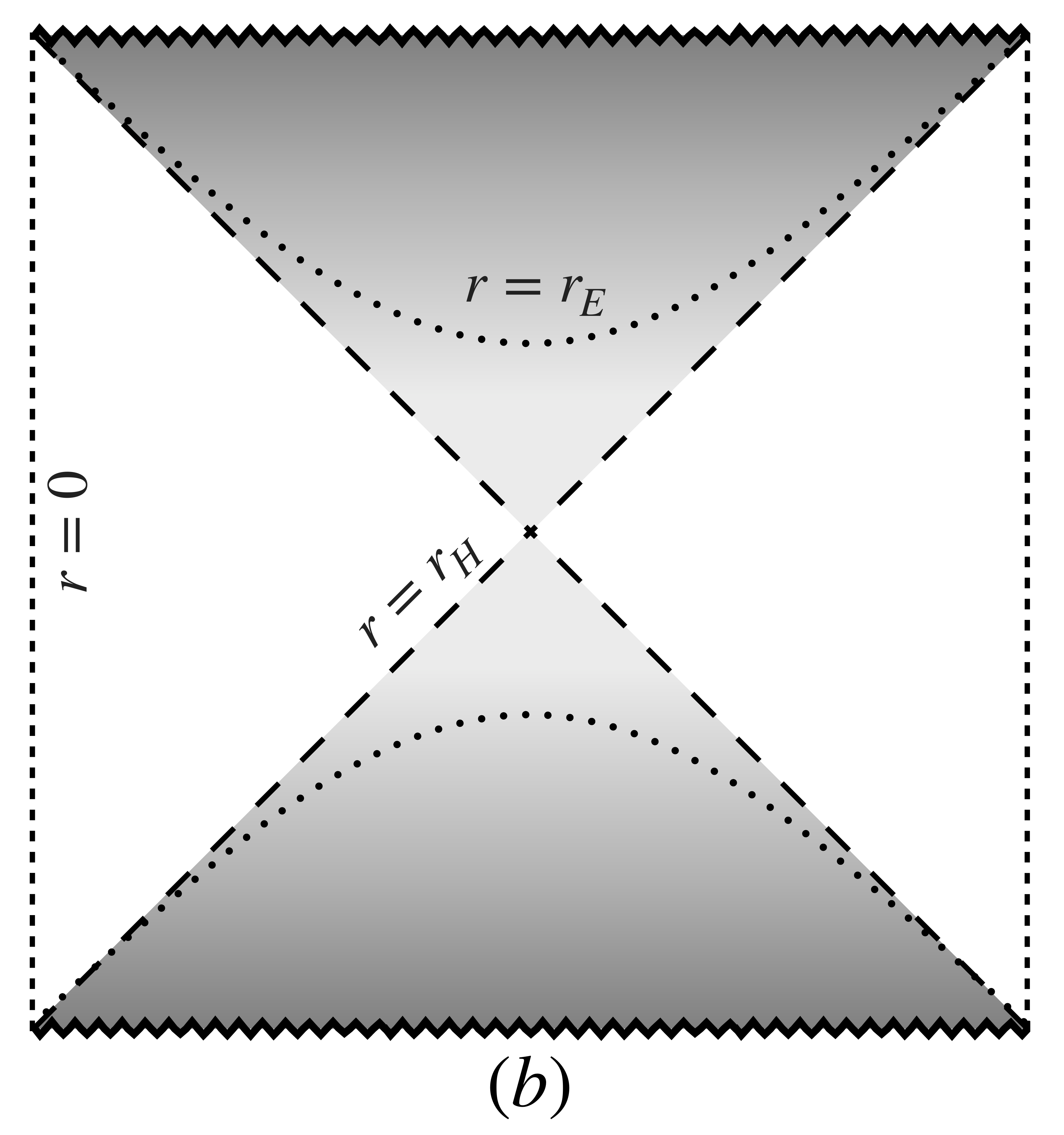}}%
\quad
{\includegraphics[width=0.31\linewidth]{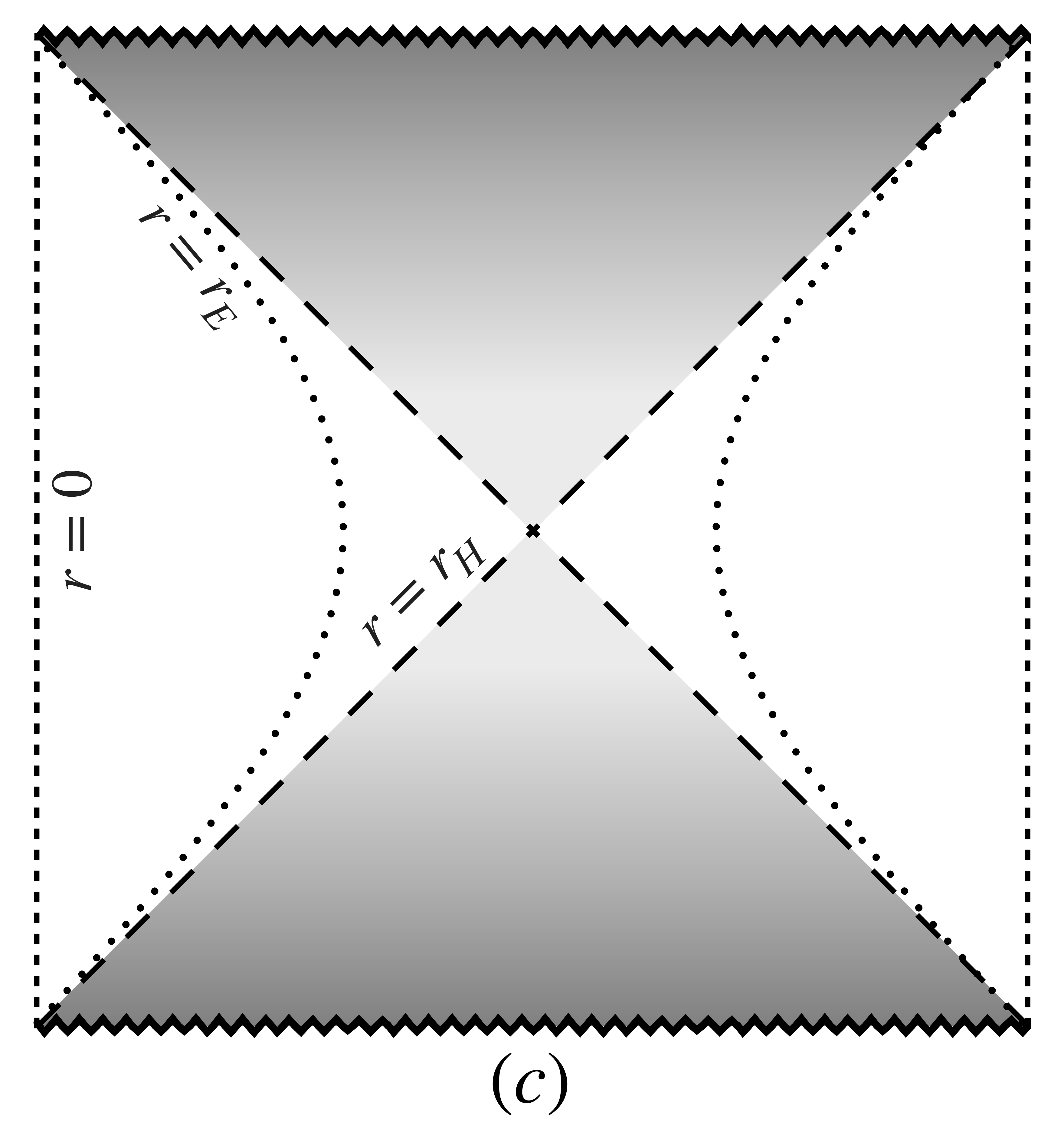}}%
  \caption{Fully extended Carter-Penrose diagrams for the coordinate neighbourhood \eqref{metricansatz}:
  shaded areas represent trapped regions, i.e. where $\theta^n\theta^l>0$,
  while the unshaded area corresponds to normal regions with $\theta^n\theta^l<0$. The dashed diagonals
  stand for the horizons, the smaller dashed lines on the left and right denote regular origins. Diagram $(a)$
  shows de Sitter asymptotics of an expanding universe with de Sitter infinity marked by the bold line on
  top and bottom.  
  Scenarios $(b)$ and $(c)$ exact the existence of a maximal surface ($R'(r)=0$ and
  $R''(r)<0$) that can be either spacelike, as in $(b)$, for $r_E>r_H$ or timelike, 
  as in $(c)$, for $r_E<r_H$ while $(a)$ prohibits its existence strictly.
  Universe $(b)$ is comparable to $(a)$ in the normal region but
  goes into a big crunch scenario for $r>r_E$ developing a singularity at the second zero of $R(r)$, here,
  marked as zig-zag line. Scenario $(c)$ reaches its maximal size in the
  normal region at $r=r_E$ and recollapses into a big crunch singularity.\label{CPD}}%
\end{figure}

In the following we focus only on the coordinate chart \eqref{metricansatz}, i.e. considering the left and upper
right-angled triangle of the Carter-Penrose diagrams in Figure \ref{CPD}; the full extension has been
created following \cite{Breitenlohner:2005}.
Taking our results from the previous sections we find for a scalar field and a regular origin that
just one horizon occurs because $f(r)$ develops only one zero leaving us with
three distinct cases (and their obvious delimiters, such as $r_E\equiv r_H$):
\begin{itemize}
\item[(a)] Blue region in Figs. \ref{fig:n1} and \ref{fig:n2}:
$R(r)$ is a monotonic function in $r$, i.e.
$\theta^l(r)=0$ at $r_H$ while $\theta^n(r)\neq0$ for all $r\in(0,\infty)$.
The scalar field self-coupling constant  $g$ is too weak compared to the cosmological constant $\Lambda$
such that the scalar lump's gravitational pull is insufficient to change the large-scale
behaviour, which leads to de Sitter asymptotics
and a cosmological horizon with
$\mathcal{L}_l\theta^n(r)>0$ at $r_H$ due to $f'(r)<0$ and $R'(r)>0$ at the horizon.
\item[(b)] Green region in Figs. \ref{fig:n1} and \ref{fig:n2}:
$R(r)$ has a local maximum at $r_E>r_H$ such that within the normal region ($0\le r<r_H$),
$f'(r)<0$ as well as $R'(r)>0$ leading to a similar cosmological horizon since
$\mathcal{L}_l\theta^n(r)>0$ at $r_H$ in this scenario.
Right after the horizon will still be a past-trapped region
that for $r>r_E$ becomes future trapped. Here, the modification
from the scalar lump only influences the asymptotic behaviour of the space-time
with the result that the universe forms a future singularity, i.e. $\theta^l(r)\to-\infty$
\cite{Liberati:2020},
after a finite time\footnote{As stated before, in the trapped region, $r$ becomes
a temporal coordinate due to the sign change in $f(r)$.}.
\item[(c)] Red region in Figs. \ref{fig:n1} and \ref{fig:n2}:
$R(r)$ has a local maximum at $r_E<r_H$ which causes $R'(r)<0$ at the horizon. In
this scenario, the influence of the
scalar lump strongly modifies the horizon structure such that $\mathcal{L}_l\theta^n(r)<0$ at $r_H$
which is associated with a black-hole like horizon. Therefore, the space-time
reaches its maximal extension in the normal region and by travelling further in $r$ one faces a black-hole
horizon shielding a future singularity.
\end{itemize}
%\footnote{Scenario (b) distinguishes between cases where $r_E<r_H$ and $r_E>r_H$
%which become important if one studies the detailed causal structure as in \cite{Senovilla:2007}
%or if one is interested
%in quantum effects induced by these horizons \cite{Giavoni:2020}.}.

%As Einstein's equations suggest, the interplay between the scalar lump and the cosmological constant fully determines the causal structure of space-time.
We see that the repulsive influence of the positive vacuum energy is successively reduced
when the coupling of the scalar field is increased. This leads to a flattening of $R(r)$ creating a
Nariai solution in the limiting case that the areal radius approaches a constant asymptotically. If the coupling
increases further the areal radius will develop a tipping point in the trapped region
that is associated with the equator, i.e. the maximal surface.
In this case, Penrose's singularity
theorem predicts the occurrence of a singularity. The higher the coupling compared to the vacuum energy, the closer the equator wanders towards the regular origin until it reaches the normal
region where the surprising scenario occurs in which the horizon becomes a black hole-like
horizon\footnote{The precise classification would state that with increasing coupling, the past inner
trapping horizon transforms into a future outer trapping horizon. Interestingly, both horizons are subjected
to the same sort of emissive Hawking effect \cite{Giavoni:2020}.}.

%%%%%%%%%%%%%%%%%%%%%%%%%%%%%%%%%%%%%%%%%%%%%%%%%

%%%%%%%%%%%%%%%%%%%%%%%%%%%%%%%%%%%%%%%%%%%%%%%%%

\section{Concluding remarks} \label{sec:discussion}

We have presented scalar lump solutions of general relativity in the presence of a scalar field with a potential barrier. The new feature of our analysis is the fact that the scalar lump spacetimes may be finite, as the size of transverse $2$-spheres may shrink again some distance away from the origin of the lump, thus forming either a collapsing universe or a surrounding black hole. In both cases, the static lump is shielded from the time-dependent collapsing regions by a horizon.

As is well known \cite{Jensen:1983ac,Battarra:2013rba},
in order for Coleman-de Luccia tunneling solutions to exist in a given potential, the
curvature at the top of the potential barrier must be sufficiently high. More precisely, for so-called oscillating instantons \cite{Hackworth:2004xb,Battarra:2012vu} interpolating $n$ times across the barrier to exist, the following inequality must be satisfied:
\be
-\frac{V''}{\kappa V/3}\mid_{\varphi=0} > n(n+3) \quad \to \quad -\frac{V''}{V}\mid_{\varphi=0} > \frac{8 \pi}{3}n(n+3) \pkt \label{cdl}
\ee
This can be compared with our case.
As already demonstrated in \cite{Torii:1999uv}, nontrivial solutions require the parameters in the potential to lie in certain ranges.
At the top of the potential barrier this criterion leads to the following inequality, making use of \eqref{lambdacritical},
\be
-\frac{V''}{V}\mid_{\varphi=0} = \frac{gv^2}{\frac{g}{4}v^4+\frac{\Lambda}{\kappa}}  > \frac{\lambda_{cr}v_{cr}^2}{\frac{\lambda_{cr}}{4}v_{cr}^4+\frac{1}{8\pi}} = \frac{8\pi}{3} 2n (2n+3) \pkt \label{cond}
\ee
%%%\begin{align}
%%%-\frac{V''}{V}\mid_{\varphi=0} = \frac{gv^2}{\frac{g}{4}v^4+\Lambda} & >\frac{\lambda_{cr}v_{cr}^2}{\frac{\lambda_{cr}}{4}v_{cr}^4+1}\\ & = %%%\frac{4n(2n+3)}{6-(8\pi - 1)(2n^2+3n)v_{cr}^2}\\
%%%& >\frac{2n}{3}(2n+3)
%%%\end{align}
Thus we can see that potentials that admit scalar lump solutions must be sufficiently (negatively) curved at the top. The similarity between the two criteria in Eqs. \eqref{cdl} and \eqref{cond} is striking, yet perhaps not entirely surprising as in both cases the inequality stems from small oscillations of a scalar around an extremum. Note however that oscillating instantons have $O(4)$ symmetry, while our solutions only have an $O(3)$ invariance.

The solutions that we have discussed do not actually extend to the potential minima. This implies that analogous solutions would also exist if the minima were replaced with functions that fell off towards zero away from the barrier, the only difference being that the damped oscillations behind the horizon would be replaced by a slow rolling of the scalar (and eventually a rapid divergence of the scalar as the singularity is approached). The important feature is the barrier, and together with \eqref{cond} we may thus say that these solutions exist in potentials that satisfy the criteria for not being in the swampland \cite{Garg:2018reu,Ooguri:2018wrx}.

Thus the solutions that we have presented may play a role in quantum gravity, in particular in the early universe. However, the analysis of  \cite{Torii:1999uv} also implies that scalar lump solutions are unstable, with a characteristic lifetime that corresponds to the Hubble rate implied by the potential energy. This raises the question as to what the end product of this instability might be. Since the solutions interpolate at least once between the two sides of the barrier, even if the solutions decay, they will leave behind domain wall structures in which the scalar field is located near the top of the potential. This is reminiscent of the scenario of topological inflation \cite{Vilenkin:1994pv}, since the topology of the solution guarantees the existence of field values high up on the potential. At first sight, the large curvature at the top of the barrier seems to preclude the existence of a viable inflationary region, but this question deserves further study. In particular, it would be useful to study the nonlinear evolution and the ultimate fate of the scalar lumps via numerical methods. We leave this interesting question for future work.

There exist at least two further avenues for investigation: in the case of Einstein-Yang-Mills theory \cite{Volkov:1996qj}, it was noted that fully regular solutions also exist, i.e. scalar lumps without any horizon. This was made possible by the particular structure of the potential in that theory. In the present theory, an obstruction to the existence of fully regular solutions is provided by Eq. \eqref{eq:fTaylorZeroGG}, which shows that the metric function $f$ necessarily decreases away from a regular origin in positive potentials, in combination with the fact that $f$ evolves monotonically. This makes it impossible to join two regular regions smoothly. However, this argument then also implies that fully regular solutions might exist when the potential minima are located at negative values of the potential, with the barrier protruding to positive vacuum energies. This is because, as implied by Eq. \eqref{eq:fTaylorZeroGG}, the metric function $f$ would first rise in a negative region of the potential, then as the scalar field crosses the top of the barrier both $f$ and $R$ could turn around, and beyond this turnaround essentially the mirrored evolution of the one just described could take place. Possibly such solutions only exist in potentials with specific values of the parameters, as was the case for a similar setup in Einstein-Yang-Mills theory \cite{Volkov:1996qj}. A second avenue for research is to look for solutions with two horizons, i.e. generalisations of the Schwarzschild-de Sitter solution containing scalar lumps between the horizons. Preliminary results indicate that in this case spherically symmetric solutions containing \emph{two} black holes exist. These will be presented in \cite{Bramberger:2021}.

\acknowledgments
The work of G.L. is supported in part by the Shota Rustaveli National Science Foundation of Georgia
with Grant N FR-19-8306 and by the DAAD scholarship
"Research Stays for University Academics and Scientists".
J.-L.L. gratefully acknowledges the support of the European Research Council in the form of the ERC Consolidator Grant CoG 772295 ``Qosmology''.
M.S. appreciates financial support from the Alexander-von Humboldt Foundation.
%%%%%%%%%%%%%%%%%%%%%%%%%%%%%%%%%%%%%%%%%%%%%%

%%%%%%%%%%%%%%%%%%%%%%%%%%%%%%%%%%%%%%%%%%%%%%%%%%%%%%%%%%

%%%%%%%%%%%%%%%%%%%%%%%%%%%%%%%%%%%%%%%%%%%%%%%%%%%%%%%%%%%%%%%%%%%%%%%%%%%%%%%%%%%%%%%%%%%%%%%

\begin{thebibliography}{99}
%%%

\bibitem{Martin:2013tda}
J.~Martin, C.~Ringeval and V.~Vennin,
``Encyclop\ae{}dia Inflationaris,''
Phys. Dark Univ. \textbf{5-6} (2014), 75-235
doi:10.1016/j.dark.2014.01.003
[arXiv:1303.3787 [astro-ph.CO]].

\bibitem{Lehners:2008vx}
J.~L.~Lehners,
``Ekpyrotic and Cyclic Cosmology,''
Phys. Rept. \textbf{465} (2008), 223-263
doi:10.1016/j.physrep.2008.06.001
[arXiv:0806.1245 [astro-ph]].

\bibitem{Clifton:2011jh}
T.~Clifton, P.~G.~Ferreira, A.~Padilla and C.~Skordis,
``Modified Gravity and Cosmology,''
Phys. Rept. \textbf{513} (2012), 1-189
doi:10.1016/j.physrep.2012.01.001
[arXiv:1106.2476 [astro-ph.CO]].

\bibitem{Obied:2018sgi}
G.~Obied, H.~Ooguri, L.~Spodyneiko and C.~Vafa,
``De Sitter Space and the Swampland,''
[arXiv:1806.08362 [hep-th]].

\bibitem{Garg:2018reu}
S.~K.~Garg and C.~Krishnan,
``Bounds on Slow Roll and the de Sitter Swampland,''
JHEP \textbf{11} (2019), 075
doi:10.1007/JHEP11(2019)075
[arXiv:1807.05193 [hep-th]].

\bibitem{Ooguri:2018wrx}
H.~Ooguri, E.~Palti, G.~Shiu and C.~Vafa,
``Distance and de Sitter Conjectures on the Swampland,''
Phys. Lett. B \textbf{788} (2019), 180-184
doi:10.1016/j.physletb.2018.11.018
[arXiv:1810.05506 [hep-th]].

%\cite{Torii:1999uv}
\bibitem{Torii:1999uv}
T.~Torii, K.~Maeda and M.~Narita,
``Can the cosmological constant support a scalar field?,''
Phys. Rev. D \textbf{59} (1999), 104002
doi:10.1103/PhysRevD.59.104002

%\cite{Volkov:1996qj}
\bibitem{Volkov:1996qj}
M.~Volkov, N.~Straumann, G.~V.~Lavrelashvili, M.~Heusler and O.~Brodbeck,
``Cosmological analogs of the Bartnik-McKinnon solutions,''
Phys. Rev. D \textbf{54} (1996), 7243-7251
doi:10.1103/PhysRevD.54.7243
[arXiv:hep-th/9605089 [hep-th]].

\bibitem{Landau}
L.D.~Landau, E.M.~Lifshitz, Course of Theoretical Physics Series,
Volume 2: The Classical Theory of Fields, Pergamon Press, 1971.

\bibitem{Nariai1951}
H.~Nariai,
``On a new cosmological solution of Einstein's field equations of gravitation",
Sci. Rep. Tohoku Univ. 35: 62 (1951)

%causal structure: references
%\cite{Ashtekar:2003}
\bibitem{Ashtekar:2003}
A. Ashtekar and B. Krishnan, ``Dynamical horizons and their properties",
Phys. Rev. D \textbf{68}, no. \textbf{10} (2003): 104030.

%\cite{Hayward:1994}
\bibitem{Hayward:1994}
S. A. Hayward, ``General laws of black-hole dynamics"
Phys. Rev. D \textbf{49},
 no. 12 (1994): 6467

%\cite{Ashtekar:2000}
\bibitem{Ashtekar:2000}
A. Ashtekar, C. Beetle, and S. Fairhurst,
``Mechanics of isolated horizons",
Class. Quant. Grav. \textbf{17}, no. 2 (2000)

%\cite{Liberati:2020}
\bibitem{Liberati:2020}
R. Carballo-Rubio, F. Di Filippo, S. Liberati, and M. Visser,
`` Geodesically complete black holes",
Phys. Rev. D \textbf{101}, no. 8 (2020)

%\cite{Mars:2003}
\bibitem{Mars:2003}
M. Mars and J. M. M. Senovilla,
``Trapped surfaces and symmetries",
Class. Quant. Grav. \textbf{20},
no. 24 (2003): L293

%cite{Senovilla:2007}
\bibitem{Senovilla:2007}
J. M. M. Senovilla,
``Classification of spacelike surfaces in spacetime",
Class. Quant. Grav. \textbf{24},
no. 11, (2007)

%\cite{Breitenlohner:2005}
\bibitem{Breitenlohner:2005}
P. Breitenlohner, P. Forg{\'a}cs, and D. Maison,
``Classification of static, spherically symmetric solutions of the Einstein-Yang-Mills theory with positive cosmological constant",
Comm. Math. Phys. \textbf{261},
no. 3, (2005)

%\cite{Giavoni:2020}
\bibitem{Giavoni:2020}
C. Giavoni, and M. Schneider. ``Quantum effects across dynamical horizons",
Class. Quant. Grav. \textbf{37}, no. 21 (2020): 215020.

%\cite{Jensen:1983ac}
\bibitem{Jensen:1983ac}
L.~G.~Jensen and P.~J.~Steinhardt,
``Bubble Nucleation and the {Coleman-Weinberg} Model,''
Nucl. Phys. B \textbf{237} (1984), 176-188
doi:10.1016/0550-3213(84)90021-X

%\cite{Battarra:2013rba}
\bibitem{Battarra:2013rba}
L.~Battarra, G.~Lavrelashvili and J.~L.~Lehners,
``Zoology of instanton solutions in flat potential barriers,''
Phys. Rev. D \textbf{88} (2013), 104012
doi:10.1103/PhysRevD.88.104012
[arXiv:1307.7954 [hep-th]].

%\cite{Hackworth:2004xb}
\bibitem{Hackworth:2004xb}
J.~C.~Hackworth and E.~J.~Weinberg,
``Oscillating bounce solutions and vacuum tunneling in de Sitter spacetime,''
Phys. Rev. D \textbf{71} (2005), 044014
doi:10.1103/PhysRevD.71.044014
[arXiv:hep-th/0410142 [hep-th]].
%94 citations counted in INSPIRE as of 27 Apr 2021

%\cite{Battarra:2012vu}
\bibitem{Battarra:2012vu}
L.~Battarra, G.~Lavrelashvili and J.~L.~Lehners,
``Negative Modes of Oscillating Instantons,''
Phys. Rev. D \textbf{86} (2012), 124001
doi:10.1103/PhysRevD.86.124001
[arXiv:1208.2182 [hep-th]].
%22 citations counted in INSPIRE as of 27 Apr 2021

\bibitem{Vilenkin:1994pv}
A.~Vilenkin,
``Topological inflation,''
Phys. Rev. Lett. \textbf{72} (1994), 3137-3140
doi:10.1103/PhysRevLett.72.3137
[arXiv:hep-th/9402085 [hep-th]].

%\cite{Bramberger:2021}
\bibitem{Bramberger:2021}
S.~F.~Bramberger, G.~Lavrelashvili and J.~L.~Lehners,
2021, In preparation.

\end{thebibliography}
\end{document}